\newtheorem{assumption}{Assumption}
\newtheorem{lemma}{Lemma}

\newtheorem{remark}{Remark}
\newtheorem{theorem}{Theorem}

\newcommand{\T}{\mathcal{T}}
\newcommand{\I}{\mathcal{I}}
\newcommand{\R}{\mathcal{R}}

\newcommand{\ED}{\textup{(ED)}}

\newcommand{\Gf}{\mathcal{G}_f}
\newcommand{\Gs}{\mathcal{G}_s}
\newcommand{\Df}{\mathcal{D}_f}
\newcommand{\Ds}{\mathcal{D}_s}
\newcommand{\B}{\mathcal{B}}

\newcommand{\E}{\mathcal{E}}
\newcommand{\graph}{\mathcal{G}}
\newcommand{\Ni}{\mathcal{N}(i)}
\newcommand{\labelP}{p^{(m)}_j}
\newcommand{\labelQ}{q^{(m)}_j}
\newcommand{\labelV}{v^{(m)}_i}

\newcommand{\gen}[1][]{p_{i,t}^{\texttt{gen}}}
\newcommand{\genupper}{\overline{\mathrm{p}}^{\texttt{gen}}_i}
\newcommand{\res}[1][]{\mathrm{P}_{i,t}^{\texttt{res}}}
\newcommand{\load}[1][]{\mathrm{P}^{\texttt{dem}}_{i,t}}
\newcommand{\loadq}[1][]{\mathrm{Q}^{\texttt{dem}}_{i,t}}
\newcommand{\demand}[1][]{p_{i,t}^{\texttt{dem}}}
\newcommand{\demupper}{\overline{\mathrm{p}}_i^{\texttt{dem}}}
\newcommand{\demandq}[1][]{q_{i,t}^{\texttt{dem}}}

\newcommand{\xa}{\tilde x}
\newcommand{\xb}{x^\star}

\newcommand{\rhv}{\rho^{\texttt{v}}}
\newcommand{\epsv}{\varepsilon_{\texttt{v}}}
\newcommand{\distPhi}{\overline{d}_{\Phi_i}}

\newcommand{\energyreq}{\mathrm{E}^{\texttt{dem}}_i}
\newcommand{\FDhorizon}{\T^{\texttt{dem}}_i}

\newcommand{\genreacupper}{\overline{\mathrm{q}}_i^{\texttt{gen}}}

\newcommand{\genreac}[1][]{q_{i,t}^{\texttt{gen}}}
\newcommand{\lbl}[1]

\newcommand{\Et}{e_{i,t}}
\newcommand{\Etminus}{e_{i,t-1}}
\newcommand{\efch}{\upeta_i^{\texttt{ch}}}
\newcommand{\efdis}{\upeta_i^{\texttt{dis}}}
\newcommand{\charge}[1][]{p_{i,t}^{\texttt{ch}}}
\newcommand{\discharge}[1][]{p_{i,t}^{\texttt{dis}}}
\newcommand{\chibat}{\upchi_i^{\texttt{sto}}}
\newcommand{\charupper}{\overline{\mathrm{p}}_i^{\texttt{ch}}}
\newcommand{\disupper}{\overline{\mathrm{p}}_i^{\texttt{dis}}}
\newcommand{\Eupper}{\overline{\mathrm{e}}_i}
\newcommand{\Elower}{\underline{\mathrm{e}}_i}

\newcommand{\chargeq}[1][]{q_{i,t}^{\texttt{sto}}}
\newcommand{\vplus}[1][]{v_{i,t}^{\ifstrempty{#1}{}{#1,}+}}
\newcommand{\vminus}[1][]{v_{i,t}^-}

\NewDocumentCommand{\vcopj}{o O{j}}{%
  \ensuremath{\hat{v}_{#2,t}[\IfValueT{#1}{#1}]}%
}
\newcommand{\rhov}[1][]{\rho_{i,j}^{#1}}

\newcommand{\aparB}{\overline{\mathrm{S}}_i^{\texttt{sto}}}
\newcommand{\resS}{\overline{\mathrm{S}}_i^{\texttt{res}}}   
\newcommand{\genS}{\overline{\mathrm{S}}_i^{\texttt{gen}}} 

\newcommand{\resq}[1][]{q_{i,t}^{\texttt{res}}}   
\newcommand{\chig}{\upchi_i^{\texttt{gen}}}
\newcommand{\chires}{\upchi_i^{\texttt{res}}}

\NewDocumentCommand{\lamp}{o o}{%
  \ensuremath{\lambda_{\IfValueT{#1}{#1,}t}^{\texttt{p}\IfValueT{#2}{,#2}}}%
}
\NewDocumentCommand{\lamq}{o o}{%
  \ensuremath{\lambda_{\IfValueT{#1}{#1,}t}^{\texttt{q}\IfValueT{#2}{,#2}}}%
}

\newcommand{\lampcopy}{\tilde \lambda^\texttt{p}_{h,t}}
\newcommand{\lamqcopy}{\tilde \lambda^\texttt{q}_{h,t}}

\newcommand{\LocalPriceP}{\lambda_{i,t}^{\texttt{p}}}
\newcommand{\LocalPriceQ}{\lambda_{i,t}^{\texttt{q}}}

\newcommand{\jresidualP}{\delta_{j,t}^{\texttt{p}}}
\newcommand{\jresidualQ}{\delta_{j,t}^{\texttt{q}}}
\newcommand{\ep}{\delta_{i,t}^{\texttt{p}}}
\newcommand{\eq}{\delta_{i,t}^{\texttt{q}}}

\newcommand{\epcopy}{\tilde\delta_{h,t}^{\texttt{p}}}
\newcommand{\eqcopy}{\tilde\delta_{h,t}^{\texttt{q}}}

\newcommand{\pcopy}{\tilde p_{i,t}}
\newcommand{\qcopy}{\tilde p_{i,t}}

\newcommand{\Vijt}{v^{(i)}_{j,t}}

\newcommand{\UpSlackVoltage}{v^{(i)+}_{j,t}}
\newcommand{\DownSlackVoltage}{v^{(i)-}_{j,t}}

\NewDocumentCommand{\p}{O{i}}{%
  \ensuremath{p_{#1,t}}%
}
\NewDocumentCommand{\q}{O{i}}{%
  \ensuremath{q_{#1,t}}%
}

\newcommand{\rhop}{\rho^{\texttt{p}}}
\newcommand{\rhoq}{\rho^{\texttt{q}}}
\newcommand{\epsp}{\epsilon^{\texttt{p}}}
\newcommand{\epsq}{\epsilon^{\texttt{q}}}

\newcommand{\vit}{v_{i,t}}
\newcommand{\ve}{\hat{v}_{i,t}}
\newcommand{\bp}[1][i,j]{\ensuremath{\mathrm{b}^\texttt{p}_{#1}}}
\newcommand{\bq}[1][i,j]{\ensuremath{\mathrm{b}^\texttt{q}_{#1}}}
\newcommand{\bz}{\mathrm{b}^{0}_i}
\newcommand{\Bg}[1][i,j]{\mathbf{B}_{#1}}
\newcommand{\pairs}{\mathcal{P}}
\newcommand{\Bone}{\ensuremath{\|\mathbf{B}\|_{1}}}
\newcommand{\Btwo}{\ensuremath{\|\mathbf{B}\|_{2,1}}}
\documentclass[journal,twoside]{IEEEtran}
\ifCLASSINFOpdf
   \usepackage[pdftex]{graphicx}
\else
   \usepackage[dvips]{graphicx}
\fi
\usepackage{enumitem}
\usepackage{xcolor}
\usepackage{siunitx}
\usepackage{mathrsfs}
\usepackage{amssymb}
\usepackage{upgreek}
\usepackage{amsmath, bm}
\usepackage{subcaption}
\usepackage{algorithm}
\usepackage{algpseudocodex}
\usepackage{derivative}
\usepackage{comment}
\usepackage{booktabs}
\usepackage{todonotes} 
\usepackage{mathtools}
\usepackage{xparse} 
\usepackage{cite}

\usepackage{url}

\makeatletter
\let\old@ps@headings\ps@headings
\let\old@ps@IEEEtitlepagestyle\ps@IEEEtitlepagestyle
\makeatother

\begin{document}
%\title{Voltage-Constrained Consensus Economic Dispatch for autonomous Local Energy Communities}
%\title{Consensus-based Voltage-Constrained Economic Dispatch for Self-organized Local Energy Communities}
%\title{Consensus Voltage-Constrained Economic Dispatch for Autonomous Local Energy Communities}
\title{Consensus-based Economic Dispatch for Self-organized Local Energy Communities}

%% To specify the authors when (number of affiliations <= 2)
\author{Dimitrios~Lagos, Josh~Eichman, 
        Georgios~Tsaousoglou

%\thanks{This work was supported in part by NEXUSFLEX project (Corresponding author: Dimitrios Lagos) and in part by the project BIPED, Grant agreement ID: 101139060.}
\thanks{D.~Lagos is with the Energy Systems Analytics Group, Catalonia Institute for Energy Research (IREC), 08930 Sant Adri\`a de Bes\`os, Spain, and also with the Department of Statistics and Operations Research, Universitat Polit\`ecnica de Catalunya (UPC), 08034 Barcelona, Spain (e-mail: dimitrios.lagos@upc.edu).}
\thanks{J.~Eichman is with the Energy Systems Analytics Group, IREC, 08930 Sant Adri\`a de Bes\`os, Spain.}
\thanks{G.~Tsaousoglou is with the Department of Applied Mathematics and Computer Science, Technical University of Denmark, 2800 Kongens Lyngby, Denmark.}
}

%% To specify the authors when (number of affiliations > 2)
% \author{\IEEEauthorblockN{Author n.1\IEEEauthorrefmark{1},
% Author n.2\IEEEauthorrefmark{2},
% Author n.3\IEEEauthorrefmark{3}, 
% Author n.4\IEEEauthorrefmark{3} and
% Author n.5\IEEEauthorrefmark{4}}
% \IEEEauthorblockA{\IEEEauthorrefmark{1} Department Name of Organization A\\
% Name of the organization A,
% Address A\\ Emails if wanted}
% \IEEEauthorblockA{\IEEEauthorrefmark{2} Department Name of Organization B\\
% Name of the organization B,
% Address B\\ Emails if wanted}
% \IEEEauthorblockA{\IEEEauthorrefmark{3} Department Name of Organization C\\
% Name of the organization C,
% Address C\\ Emails if wanted}
% \IEEEauthorblockA{\IEEEauthorrefmark{4}Department Name of Organization D\\
% Name of the organization D,
% Address D\\ Emails if wanted}
% }

\maketitle

\begin{abstract}
The increasing penetration of distributed energy resources (DERs) is transforming traditional distribution networks into dynamic, decentralized systems, creating a pressing need for scalable, autonomous coordination mechanisms that do not rely on centralized control or full system observability. In particular, local energy communities need to self-organize and coordinate energy schedules while respecting network constraints, especially voltage limits, despite lacking detailed knowledge of the distribution grid's topology and parameters. Existing distributed optimization approaches either neglect such physical constraints or assume perfect knowledge of the network model. This paper addresses the Voltage-Constrained Economic Dispatch (VCED) problem in a fully decentralized, consensus-based manner, without any coordinating entity. We use sparse-group LASSO regression to learn a linear surrogate of nodal voltages from local power and voltage measurements alone. The surrogate serves a dual purpose: it enters each agent's local problem as a voltage penalty and induces the communication graph over which the agents reach consensus. 
We provide a theoretical bound on the optimality loss caused by the surrogate voltage estimator inaccuracy.
Our numerical results, on an IEEE-33 network, show that the scheme reduces the share of bus-timesteps voltage violations from 30\% to below 3\%, while staying within 1\% of a centralized benchmark computed with full network knowledge. We further quantify how the sparsity of the communication graph affects both the optimality gap and the residual violations.
%the proposed scheme reduces the share of bus-timesteps outside the admissible voltage range from 30\% to below 3\%, across communication graphs spanning a wide range of sparsity levels.
\end{abstract}


\section{Introduction}

\subsection{Background and Motivation}

Power systems are undergoing a structural transformation where rooftop PV, home batteries, electric vehicles, and flexible loads are turning millions of individual consumers into active, flexible nodes. As such, when prosumers participate in electricity markets (namely through aggregators or local energy communities) the safe operation of the power distribution network is challenged. Power injections and withdrawals that violate voltage limits trigger intervention by the Distribution System Operator (DSO), which is responsible for restoring feasibility. In practice, the DSO typically does that through coarse, horizontal corrective actions, without the prosumers' economics in mind. 
This creates a natural incentive for prosumers to account for network constraints \textit{proactively} in their own decisions, rather than leaving safety to (typically inefficient) downstream corrections. 

Being proactive about network safety is difficult in practice, because it requires a model of the network, its topology, and its parameters; information that is available exclusively to the DSO (or, often, not even to the DSO). 
However, what prosumers do have is access to locally measured data, namely their power profiles and corresponding local voltages.
This enables the development of data-driven, topology-agnostic surrogates, learned purely from measurements, that let prosumers anticipate the impact of their actions on voltages.

Equipped with such a surrogate, prosumers face an (approximate) instance of the system's Optimal Power Flow (OPF) problem. 
Naturally, as the number of active nodes increases, scalability becomes a challenge.
This is well-recognized in the literature, which has responded with a rich body of work on distributed OPF, decomposing the problem via dual decomposition, ADMM, and related methods. 
However, such algorithms still lean on a centralized step: a coordinator that aggregates information from all agents and broadcasts a result back (e.g., nodal prices / multipliers). This step necessarily assumes a stable, fully-connected path to every node and a \textit{static} system. In the face of a system that constantly changes (topology is reconfigured, new assets connect at nodes that were previously inactive, communication links drop or appear, etc.), plain distributed OPF solutions fail a reality check.

The ever-evolving nature of power distribution systems calls for coordination schemes that do not depend on a fixed communication structure: a protocol by which thousands of agents reliably reach a globally consistent, efficient, and safe operating point, hour after hour, despite changing conditions and system structure.
This is what consensus-based algorithms provide. By replacing the single centralized aggregation step with repeated local exchanges among neighbors, they guarantee convergence to the correct global solution under time-varying communication graphs.

In this paper, we design a consensus-based algorithm to coordinate the energy management decisions of prosumers without assuming access to network information.
We argue that such a data-driven and fully decentralized scheme offers a pathway towards a genuinely self-organized and bottom-up adaptive power system.

\subsection{Related Work}

%\todo[inline]{Review and revise this section: Structure the related work section in two parts, following the aspects motivated above: one about data-driven OPF and one about distributed OPF (ADMM etc). Note that in this subsection, you don't talk about our paper at all - only about what others did and didn't do.}

This section reviews the two categories of work most relevant to our approach: explicit OPF formulations (focusing on their distributed variants) and data-driven OPF frameworks. The former enforce network feasibility through explicit, most often relaxed versions of the grid model, while the latter use data provided by the network to reconstruct missing information in the proposed models.

%A significant body of work spans the classic EDP–OPF spectrum, much of it through the distributed frameworks that are of primary interest here. Methods targeting the EDP typically coordinate active power generation across agents while 
Neglecting network constraints keeps the problem convex and amenable to standard consensus algorithms (as presented in \cite{Kar_Hug_2012, Hug_2015, Papadaskalopoulos_2021} and more recently \cite{Wu_2026}). In contrast, OPF formulations explicitly incorporate the AC power flow equations and voltage limits, yielding nonconvex problems that are commonly handled through convex relaxations. A prominent case is the second-order conic relaxation \cite{Low_OPF_relax_2014}, while distributed variants thereof were presented in \cite{Sleiman_2019_OPF_ADMM, Cheng_2021_collab_opf}. Another case is the linearized approximation of \cite{Baran_1989_LinDistFlow} and its decentralized variant \cite{dobbe2017fully}. 
%While the works reviewed above occupy different points along the EDP–OPF spectrum, trading exactness against tractability, a common feature limits their real-world applicability: 
These models require prior knowledge of the physical system model and its topology, which, in our context, limits their applicability.

Data-driven surrogate models offer a natural way to bypass this requirement, and a substantial body of work has developed voltage–power mappings under this premise. These efforts fall into two approaches. The first is completely data-driven, assuming zero information about the grid, while the second assumes that a grid model is generally available but part of the information is missing.

The literature in the first category comprises linear and low-order surrogates e.g., \cite{Zhigang_Log, Li_2022_proceedings}, regression-based 
models (e.g. Bayesian and LASSO) \cite{Guoan_2024, Liu_2019}, piecewise-linear approaches based on operating-point classification~\cite{Long_2024}, and 
feature-augmented linear/quadratic compositions~\cite{Zhang_2024, Guo_2022}. These yield closed-form expressions that can in principle be embedded directly as affine or convex quadratic constraints in an optimization problem. 

In the second category, 
%Data-Driven OPF models, solve the OPF assuming that a part of information of the grid's model is missing. For instance, 
\cite{Li_2023_Data_OPF}
%, having access to the grid's topology, 
combines Support Vector Regression with Ridge Regression to incorporate Kirchhoff's law into a linearized network model.
The authors in \cite{Liu_2024_DATA_OPF_UaM} build a convex uncertainty-aware OPF model for assessing the static voltage stability margin.
%, under the same assumption. 
On the other hand, \cite{Model_Free_OPF_Simonovska_2025} proposed a neural-network-based surrogate. For our purposes, however, it is opaque
%achieve high accuracy but, due to their opaque 
structure, does not allow for incorporating it as hard or soft constraints in a tractable distributed optimization scheme.
%without significant relaxation. 
%Closest in spirit is the 
A model-free extremum-seeking scheme is proposed in \cite{Arnold_2018}, which achieves voltage regulation without a grid model. However, it does so through real-time probing rather than optimization, and never recovers the inter-bus structure.
%that our learned sparse surrogate exploits.

\subsection{Research Gap \& Contributions}

The literature review above highlights an important research gap. While distributed OPF algorithms are abundantly present in the literature, fully decentralized (consensus-based) algorithms are barely present. The few studies that have pursued this (\cite{Hug_2015, Papadaskalopoulos_2021, Wu_2026}) have restricted themselves to the plain economic dispatch problem, without consideration of network constraints. This is not a coincidence, since integrating voltage considerations into the consensus algorithm, especially when access to the grid's topology is absent, bears two major challenges.
%not a simple extension,
The first challenge refers to obtaining a voltage mapping, in a form that can be integrated into the constraints of an optimization problem, without using topology information.
%, suitable as an optimization input—tractable and requiring no prior topology knowledge—is not straightforward. 
The second is about the consensus algorithm itself, which relies on a communication graph and the associated doubly stochastic matrix (\cite{Kar_Hug_2012, Hug_2015, Papadaskalopoulos_2021, Wu_2026}) to diffuse the Lagrange multipliers and the locally estimated power residuals: How such a graph should be constructed is not addressed in the reviewed literature.

Our technical contribution precisely combines these two elements: we train a sparse surrogate that not only enforces voltage feasibility but also defines the communication graph, connecting two buses whenever one has a significant data-driven (P, Q)–voltage sensitivity with respect to the other.
Thereby, connections are shaped in a data-driven manner, without knowledge of the grid's topology.
%not only by the grid topology but also by each bus's activity. 
Given that the resulting graph has arbitrary structure, we make no assumptions on the communication matrix being doubly stochastic (which is a requirement for the consensus algorithm's convergence).
%, unlike the regular graphs assumed by previous approaches, 
Instead, we ensure convergence by applying a classical Sinkhorn–Knopp algorithm to restore double stochasticity ex post.
In summary, the paper's contributions read as follows:

\begin{itemize}
    %\item Upgrade the problem to Voltage-Aware Economic Dispatch without access to the topology and parameters of the distribution network, making the approach more broadly applicable in practice.
    %\item Improve the state-of-the-art algorithm by enabling the establishment of a communication graph with arbitrary topology.
    \item We design a consensus-based algorithm for fully decentralized coordination of active prosumers, proactively accounting for voltage safety via a data-driven surrogate without access to network topology and information.
    \item We prescribe an algorithm that derives the agents' communication graph from the same surrogate, connecting two buses whenever one has a significant data-driven $(P, Q)$--voltage sensitivity with respect to the other, and we show how a doubly stochastic communication matrix is obtained for the arbitrary graph that results.
    \item We characterize the fixed points of the proposed scheme, showing that they solve the community dispatch problem at a common price across agents, and we evaluate convergence, voltage feasibility and cost across a range of graph sparsity levels and DER penetration scenarios.
\end{itemize}

%Minor:
%\begin{itemize}
%    \item Demonstrate that the proposed distributed algorithm converges within a short computation time, even when run on limited computational resources, using an 8-time step, 1-hour-resolution (8-hour) horizon representative of intra-day scheduling. This suggests the framework's suitability for a range of horizon lengths and time resolutions, from intra-day to shorter, more granular scheduling windows.
 %   \item Expansion of the problem via consideration of reactive power and more detailed modeling of assets, especially power generation.
%\end{itemize}

%\subsection{Structure of the paper}

The remainder of this paper is organized as follows. Section~\ref{Sec:Problem_Form} presents the problem formulation. 
%of the Voltage-Aware EDP, considering reactive power and the modeled assets. 
Section~\ref{Sec:Method} describes the proposed voltage estimation approach
%, based on Sparse Grouped LASSO (SGL) regression, 
and its use in selecting the communication graph, along with the proposed consensus-based approach.
%used to incorporate it into the EDP. 
Section~\ref{Sec:Results} presents the evaluation scenario and numerical results, assessing estimation accuracy, the accuracy-sparsity trade-off, voltage constraint satisfaction, computational performance, and a sensitivity analysis. Finally, Section~\ref{Sec:Conclusions} concludes the paper.
%and outlines directions for future work.

\section{Problem Formulation} \label{Sec:Problem_Form}
%In this section we formulate the problem addressed in this work. We first introduce the notation and the main quantities used throughout the formulation. We then present the voltage-aware joint active–reactive power economic dispatch problem with flexible demand.

%\subsection{Abbreviations and Acronyms}
%This subsection summarizes the main notation, indices, sets, parameters, and variables used throughout the paper. 
We consider the operation of a set of distributed energy resources (DERs), connected to a power distribution network,
%, including flexible and inflexible generation, flexible and inflexible demand, and storage.
%the operation of a 
%set of distributed energy resources (DERs), connected to 
%a power distribution network and 
within a horizon $\T$ of discrete time steps $t \in \T$ of equal duration $\tau$.
%and iterative coordination rounds $r \in R$.
%The network comprises a set $\I$ of nodes.
%Each node $i \in \I$ features
The DERs under a node $i \in \I$ of the network are managed by an agent, and we thereby use the terms ``node'' and ``agent'' interchangeably. For ease of presentation, we consider one agent per node, without loss of generality.

%The subsets $\Gf \subseteq I$ and $\Df \subseteq I$ represent agents with dispatchable generation and flexible demand, respectively, while $\Gs \subseteq I$ and $\Ds \subseteq I$ denote agents with inflexible generation and demand. The subset $\B\subseteq \I$ represents agents with energy storage unit. The temporal resolution of the model is given by $\tau$.

%multiple types of DERs among: a set $\G$ dispatchable generators 
%(DGs)
%, flexible demands (FDs), energy storage units (ES) and inflexible generators (IGs) or demands (IDs).

\subsection{DER Models}

Regarding a node's generation capabilities, let $\gen$ denote the active power generation of node $i$ at time step $t$,
%\textcolor{blue}{The generation cost is given by an increasing convex function $f_i(\gen)$,}
which is subject to a capacity limit:
\begin{equation}
    0 \leq \gen \leq \genupper, \quad \forall t \in \T. \label{eq:gen_lims}
\end{equation}
Inflexible generators (e.g., renewables), instead inject a fixed, non-dispatchable output $\res$. Both flexible and inflexible generators can also generate reactive power indicated by $\genreac$ and $\resq$ respectively:
\begin{align}
&-\genreacupper \leq \genreac \leq \genreacupper, \ \forall t \in \T \label{eq:gen_q_lims}\\
&-\genreacupper \leq \resq \leq \genreacupper, \ \forall t \in \T. \label{eq:res_q_lims}
\end{align}
The reactive power output is further limited by the apparent-power rating $\overline{\mathrm{S}}_i^{(\cdot)}$ of each generator's inverter and by a minimum power factor $\mathrm{pf}^{(\cdot)}_i$, encoded by
$\chi^{(\cdot)}_i = \tan\!\left(\arccos \mathrm{pf}^{(\cdot)}_i\right)$. The coupling between active
and reactive components is modeled after \cite{Wang_2025, daCosta_2024} and, for flexible generators, reads:
\begin{align}
&-\chig \, \gen
  \leq \genreac
   \leq \chig \, \gen,
  && \forall t \in \T \label{eq:chi_couple_gen} \\
&\left(\gen\right)^{2} + \left(\genreac\right)^{2}
  \leq \left(\genS\right)^{2},
  && \forall t \in \T \label{eq:aparent_couple_gen}
\end{align}
while, for renewable ones, reads:
\begin{align}
&-\chires \, \res
  \leq \resq
   \leq \chires\, \res,
  && \forall t \in \T \label{eq:chi_couple_res}\\
&\left(\res\right)^{2} + \left(\resq\right)^{2}
  \leq \left(\resS\right)^{2},
  && \forall t \in \T. \label{eq:aparent_couple_res}
\end{align}

The consumption of the node's flexible demand is denoted by $\demand$, and is subject to an upper bound $\demupper$:
\begin{equation}
    0 \leq \demand \leq \demupper, \quad \forall t \in \T. \label{eq:demand_lim}
\end{equation}
Flexible demand has an energy requirement $\energyreq$ to be satisfied within a given horizon $\FDhorizon$:
\begin{equation}
    \sum_{t \in \FDhorizon}\demand \tau = \energyreq,
    \label{eq:FDhorizon_demand}
\end{equation}
while assumed to feature a
fixed power factor $\mathrm{pf}^{\texttt{dem}}_i$, encoded by
$\uppsi_i = \tan\left(\arccos \mathrm{pf}^{\texttt{dem}}_i\right)$. The reactive demand $\demandq$ is coupled to the controllable active demand
$\demand$ through
\begin{align}
\demandq = \uppsi_i\, \demand,
  \quad \forall t \in \T. \label{eq:demand_q_couple}
\end{align}
Finally, a node's inflexible active and reactive demand is denoted by $\load$ and $\loadq$ respectively.

Regarding storage, $\Et$ is the state of charge at time step $t$, updated from the charging and discharging powers $\charge$ and $\discharge$
according to

\begin{align}
\Et = \Etminus + \efch \, \charge
  - \discharge / \efdis,
  \quad \forall t \in \T \label{eq:bat_balance}
\end{align}
where $\efch$ and $\efdis$ are the charging and
discharging efficiencies. The state of charge is
bounded, and the charging and discharging powers are limited by:
\begin{align}
\Elower &\leq \Et \leq \Eupper,
  && \forall t \in \T \label{eq:Bat_State_lims} \\
0 &\leq \charge \leq \charupper,
  && \forall t \in \T \\
0 &\leq \discharge \leq \disupper,
  && \forall t \in \T. \label{eq:dis_lims}
\end{align}
To avoid net energy drift over the horizon, the final state of charge is constrained to be at least equal to the initial one:
\begin{align}
e_{i,|\mathcal{T}|} \geq e_{i,0}.
\end{align}
Finally, the reactive power $\chargeq$ of the storage inverter is limited by
its minimum power factor, encoded by $\chibat$, and coupled to the net
active power through the apparent-power rating $\aparB$:
\begin{align}
&-\chibat\, \charupper
  \leq \chargeq \leq \chibat\, \charupper,
  && \forall t \in \T \label{ineq:battery_reactive}\\
 &\left(\discharge - \charge \right)^{2}
  + \left( \chargeq\right)^{2}
 \leq \left(\aparB\right)^{2},
  && \forall t \in \T. \label{eq:bat_aparent}
\end{align}

\subsection{DER Community Economic Dispatch}

We consider the case where the agents (prosumers) organize and coordinate their operational decisions, i.e. forming a so-called local energy \textit{community}.
Renewable generation, battery power, and flexible demand are regarded as zero marginal cost resources, while the cost of active power generation from conventional generators is given by an increasing convex function $f_i(\gen)$. The community can also draw (inject) energy from (to) the upstream grid at a time-of-use buy (sell) price; this is modeled through a special generator at the slack node with linear (positive and negative) cost function $f_0(p^{\texttt{gen}}_{0, t})$.

The net active power generation  $p_{i,t}$ of an agent (node) $i \in \I$ (positive for net generation and negative for net consumption) is given by:
\begin{equation}
p_{i,t} = \load
  +  \demand
  - \gen 
- \res
  - \discharge + \charge, \quad \forall t \in \T,
\label{eq:agent_active_power_balance}    
\end{equation}
and, analogously, the net reactive power generation is:
\begin{equation}
q_{i,t} = \loadq
  + \demandq
  - \genreac - \resq
  - \chargeq \quad \forall t \in \T.
  \label{eq:agent_reactive_power_balance}
\end{equation}

Each agent only has access to local measurements of active and reactive power, and nodal voltage $\vit$. These are linked through a mapping
\begin{align}
\vit = \Phi_i \Big( (p_{j,t})_{j \in \I}, (q_{j,t})_{j \in \I}\Big),
  \quad \forall t \in \T,
  \label{eq:voltage_equation}
\end{align}
which is unknown to the agents, since they have no information on the underlying power distribution network.

The community's goal is to minimize the agents' aggregated energy cost while proactively ensuring that the joint operational decisions do not result in voltage limit violations.
This goal is formalized as the so-called Voltage Constrained Economic Dispatch (VCED) problem:
\begin{align}
    \min_{(x_i)_{i \in \I}} &~\left\{\sum_{t \in \T} \sum_{i \in \I} f_i(\gen) \right\} \tag{VCED} \label{centralized} \\
     \text{s.t.:} 
            &~ \eqref{eq:gen_lims} - \eqref{eq:voltage_equation}, \quad &\forall i \in \I, \notag \\
     &~ \sum_{i \in \I} p_{i,t} = 0 \;:\; (\lamp), \quad &\forall t \in \T \label{eq:active_power_balance}\\
      &~ \sum_{i \in \I} q_{i,t} = 0  \;:\; (\lamq), \quad &\forall t \in \T \label{eq:reactive_power_balance}\\
       &~ \underline{V} \leq \vit \leq \overline{V}, \quad &\forall i \in \I, t \in \T, \label{eq:voltage_limits}
\end{align}
where the decision variables $x_i$ are
\begin{equation*}
x_i = \big(p^{\texttt{gen/dem/ch/dis}}_{i,t}, q^{\texttt{gen/res/dem/sto}}_{i,t}, p_{i,t}, q_{i,t}, \vit \big)_{t \in \T},
\end{equation*}
and $\lamp, \lamq$ denote the dual variables (Lagrange multipliers) of the community's active and reactive power balance constraints.
Also note that, by relaxing the constraints \eqref{eq:voltage_equation} and \eqref{eq:voltage_limits} related to voltages, problem \eqref{centralized} reduces to the plain Economic Dispatch (ED) problem, which we shall use as a benchmark for comparison.

In line with the adaptive, self-organized system motivated in the introduction, our goal is to approach a solution to problem \eqref{centralized} through a fully decentralized dispatch scheme. 
An additional challenge, however, is that the network information (and hence the mapping $\Phi_i$ of Eq. \eqref{eq:voltage_equation}) is not known to the agents' community, which renders it impossible to evaluate Eq. \eqref{eq:voltage_equation} directly.
%Moreover, 
The next section addresses both of these aspects.

\section{Consensus-Based Solution Approach} 
\label{Sec:Method}

%\textcolor{blue}{Begin with the Consensus, using a graph and a surrogate as placeholders. At the end of the subsection, say that the instantiation of the graph needs to address a trade-off: on one hand, it needs to be as sparse as possible to minimize communication overhead and privacy leakage (is there another reason?), while on the other hand it needs to remain doubly stochastic so that you get convergence, AND connected enough so that it supports a sufficiently accurate surrogate model for the voltage estimator. Then, subsection B will present the surrogate and Algorithm 1, towards addressing these aspects.}

In this section, we present our proposed approach towards a decentralized solution to problem \eqref{centralized}.
To that end, we first present a consensus algorithm (Subsection~\ref{sec:consensus}) through which the agents can coordinate their actions using some communication graph and some voltage estimation model. Then we actually instantiate these two elements.
In Subsection~\ref{sec:surrogate} we present a method for constructing a voltage estimator, based solely on available measurements, while in Subsection~\ref{sec:graph} we present our proposed methodology towards instantiating the consensus algorithm's communication graph, which needs to address a trade-off: on one hand it needs to be as sparse as possible to minimize communication overhead and privacy leakage, while on the other hand it needs to fulfill certain properties so that convergence of the consensus algorithm is retained. Moreover, the edges need to be chosen intentionally such that the voltage estimator of Subsection~\ref{sec:surrogate} remains sufficiently accurate.

%propose a method to learn a surrogate model for the power-to-voltage mapping $\Phi(\cdot)$, trained through local measurements (subsection \ref{sec:surrogate}), and we then integrate it into a consensus algorithm (subsection \ref{sec:consensus}) that reconciles a solution in a decentralized way.

%we first propose a method to estimate the voltage, which in turn induces a communication graph. We then integrate the voltage estimator into the consensus-based EDP, resulting in a surrogate of the OPF that we refer to as the Voltage-Surrogate OPF (VSOPF).

\subsection{Consensus-Based Voltage-Constrained Economic Dispatch} \label{sec:consensus}

In this subsection, we present the consensus algorithm for problem \eqref{centralized}, considering
\begin{itemize}
    \item {a given undirected communication graph} $\graph := (\I, \E)$, where $\E \subseteq \I^2$ is the set of edges $(i,j)$,
    \item a corresponding doubly stochastic communication matrix $\mathbf{W}$, and
    \item a given, additively separable across agents, voltage surrogate model $\phi(\cdot)$:
\begin{equation}
    \vit = \phi_i \Big( (p_{j,t})_{j \in \Ni}, (q_{j,t})_{j \in \Ni}\Big),
  \label{eq:abstract_voltage_estimator}
\end{equation}
which estimates the voltage of node $i$ using the active and reactive power of $i$'s neighbors as input. 
\end{itemize}
All three elements ($\graph$, $\mathbf{W}$, and $\phi(\cdot)$) are kept abstract for now and instantiated in the following subsections. The neighborhood of $i$ is defined as $\Ni := \{j \in \I: (i,j) \in \E \}$ and it includes $i$ by convention i.e. $(i,i)\in\E,~\forall i \in \I$. 
%\textcolor{blue}{In our distributed framework, also, each agent $i$ has access to is corresponding neighborhood , $\Ni$, and its corresponding row, $W_i$, of the communication matrix.}

%In this section, we present the Consensus-Based method proposed and studied in the bibliography, similarly to \cite{Papadaskalopoulos_2021} and \cite{Wu_2026}, adding reactive power components and including surrogate voltage constraints for penalizing predicted voltage violations. 
%There are four main steps that constitute the full distribution of the algorithm in a consensus-based manner, which are the initialization, Lagrange multipliers update, obtaining the distributed solution, and sharing of the necessary information between the agents. First, we describe how the power balance is handled in the literature and extend it to include the reactive-power component, and later we incorporate the voltage constraints into the decentralized framework, yielding a distributed method that solves a relaxation of the problem in Eq.~\eqref{eq:cost_function_simple}.

%Towards a decentralized scheme, each agent updates its own local decisions $x_i[k]$ across a series of algorithm iterations $k$.

In problem \eqref{centralized}, the objective function and the local constraints \eqref{eq:gen_lims}--\eqref{eq:agent_reactive_power_balance} are directly decomposable into per-agent local problems.
However, there are two distinct types of coupling constraints: 
the community's active and reactive power balance constraints \eqref{eq:active_power_balance}--\eqref{eq:reactive_power_balance}, and the surrogate voltage estimation constraints \eqref{eq:abstract_voltage_estimator} which replace Eq. \eqref{eq:voltage_equation} in problem \eqref{centralized}.
To ensure global consistency in a decentralized way, agents coordinate their local decisions $x_i[k]$ across a series of algorithm iterations $k$. At each iteration, each agent updates its local decisions $x_i[k]$ accounting for its local objective constraints, as well as for both types of coupling constraints. In what follows, we first present how the coupling constraints are represented locally by agent $i$, and we then use these formulations to present the consensus algorithm.

\subsubsection{Power Balance Coupling Constraints}

Towards accounting for the community's power balance equations \eqref{eq:active_power_balance}--\eqref{eq:reactive_power_balance}, each agent maintains local proxies $\LocalPriceP, \LocalPriceQ$ of the Lagrange multipliers $\lamp, \lamq$ corresponding to \eqref{eq:active_power_balance}--\eqref{eq:reactive_power_balance}.
%Each agent $i$, updates t
At iteration $k$, these proxies are updated as
\begin{align}
    \LocalPriceP[k] =  \sum_{j \in \Ni}W_{i,j} \lambda^{\texttt{p}}_{j,t}[k-1] + \epsp \ep[k-1]  
    \label{eq:Plambda_update}\\
    \LocalPriceQ[k] =  \sum_{j \in \Ni}W_{i,j} \lambda^{\texttt{q}}_{j,t}[k-1] + \epsq \eq[k-1]
    %\delta^{\texttt{q}}_{i,t}[k-1],  
    \label{eq:Qlambda_update}
\end{align}
where $\epsilon$ is a step size parameter, $W_{i,j}$ is the weight that $i$ puts on $j$'s corresponding local estimate/ proxy (i.e., the $(i,j)$ element of $\mathbf{W}$), and $\ep, \eq$ are the power balance residuals as estimated locally by agent $i$. These residuals are linked to $\p$ and $\q$ by:
\begin{align}
    \ep &= \sum_{j \in \Ni} W_{i,j}\, \jresidualP[k-1] + \p - \p[i][k-1],
    \label{eq:ResidualP}\\
    \eq &= \sum_{j \in \Ni} W_{i,j}\, \jresidualQ[k-1] + \q - \q[i][k-1],
    \label{eq:ResidualQ}
\end{align}
where $\jresidualP[k-1]$ and $\jresidualQ[k-1]$ are the active and reactive residual values that agent $i$ received from its neighbors $j \in \Ni$ at iteration $k-1$, including its own, and $\p[i][k-1], \q[i][k-1]$ are its net injections at that iteration.

Once $i$ has updated $\LocalPriceP[k], \LocalPriceQ[k]$ and its expressions of $\ep, \eq$ (cf Eqs. \eqref{eq:ResidualP}, \eqref{eq:ResidualQ}), it updates its local augmented Lagrangian function:
\begin{align}
\mathcal{L}_i(x_i) =  \sum_{t \in \mathcal{T}} \Bigg[ f_i(\gen) + \LocalPriceP[k] \ep + \LocalPriceQ[k]  \eq + \notag\\ 
\frac{\rhop}{2} (\ep)^2 + \frac{\rhoq}{2}(\eq)^2 \Bigg].
\label{eq:aug_lagrang_indiv}
\end{align}
%
%where
%\begin{align}
 %   \LocalResidualP =& ~p_{i,t} - \load -  \demand + \gen  + \res + \discharge - \charge \\
  %\LocalResidualQ =& ~q_{i,t} - \loadq - \demandq + \genreac + \resq + \chargeq.
%\end{align}
% for the active and reactive power residuals of its own neighborhood, with positive $\ep$ (respectively $\eq$) representing active (respectively, reactive) power excess and negative representing deficit.
%, as in:
%updates $x_i$ and, consequently, $\ep[k], \eq[k]$. Finally, $\ep[k-1]$ (respectively $\eq[k-1]$), which is $i$'s last estimate of the active (respectively reactive) power residual, in turn, are updated as:
%based on the information about active and reactive power residuals $\ep$ and $\eq$ that the agent gets from its neighbors:
\begin{comment}
\begin{align}
    \ep[k] = \sum_{j \in \Ni} W_{i,j} \jresidualP[k-1] + p_{i,t}[k] - p_{i,t}[k-1] \label{eq:ResidualP}\\
    \eq[k] = \sum_{j \in \Ni} W_{i,j} \jresidualQ[k-1] + q_{i,t}[k] - q_{i,t}[k-1]. \label{eq:ResidualQ}
\end{align}
\end{comment}

\subsubsection{Voltage Constraints}

Towards accounting for the voltage limits \eqref{eq:voltage_limits}, each agent $i$ needs to account for how its own decisions $x_i[k]$ affect the resulting voltages of its neighbors. 
To this end, $i$ can use the voltage estimator model of Eq. \eqref{eq:abstract_voltage_estimator} to account for a local estimation $\Vijt$ of $j$'s voltage as:
\begin{equation}
 \Vijt =  \Vijt[k-1] +  \phi_j ( p_{i,t}, q_{i,t}) - \phi_j ( p_{i,t}[k-1], q_{i,t}[k-1]). \label{eq:voltage_individual_estimation}    
\end{equation}
Let $\UpSlackVoltage, \DownSlackVoltage$ denote the auxiliary variables representing the slack by which $i$'s estimate of $j$'s voltage violates the voltage limits of Eq. \eqref{eq:voltage_limits}, as in:
\begin{align}
    \UpSlackVoltage \geq \Vijt - \overline{V} \label{eq:UpSlack} \\
    \DownSlackVoltage \geq \underline{V} - \Vijt. \label{eq:DownSlack}
\end{align}
Agent $i$ accounts for the voltage limits of its neighbors through a function $\mathcal{V}(x_i)$ that penalizes the violation slacks, as well as the difference between $i$'s last received update $v_{j,t}[k-1]$ of $j$'s voltage, and $i$'s current estimate thereof, as in:
\begin{multline}
    \mathcal{V}_i\left(x_i, \UpSlackVoltage, \DownSlackVoltage, \Vijt \right) = \\
    \sum_{j \in \Ni} \rhv \bigg((\UpSlackVoltage)^2 +
    (\DownSlackVoltage)^2 \bigg) 
    + \\ \sum_{j \in \Ni\setminus \{i\}} \rho^{\delta\texttt{v}} \big(\Vijt - v_{j,t}[k-1] \big)^2. 
    \label{eq:F_indiv_voltage}
\end{multline}

\subsubsection{Consensus Algorithm}
Based on the functions that represent the coupling constraints as defined in the previous two subsections, agent $i$'s local optimization problem \eqref{LocalProblem} to be solved at each iteration $k$ reads:
\begin{align}
    \min_{\mathcal{X}_i} &\left\{ \mathcal{L}_i(x_i) + \mathcal{V}_i\left(x_i, \UpSlackVoltage, \DownSlackVoltage, \Vijt \right) \right\} \tag{Local} \label{LocalProblem}\\
    \text{s.t.}~ &\eqref{eq:gen_lims}-\eqref{eq:agent_reactive_power_balance}, \notag\\
    %&\eqref{eq:aug_lagrang_indiv} \notag\\
    & \eqref{eq:ResidualP}-\eqref{eq:ResidualQ},\quad \forall t \in \T  \notag \\
    & \eqref{eq:voltage_individual_estimation}-\eqref{eq:DownSlack}, \quad \forall j \in \Ni, t \in \T.  \notag
    %&\eqref{eq:F_indiv_voltage}.\notag
\end{align}

\begin{algorithm}[t]
\caption{Consensus Algorithm}
\label{alg:ConsensusAlgorithm}
\begin{algorithmic}[1]
\Require $\Ni, W_i, \phi_i(\cdot), \epsp, \epsq, \rhop, \rhoq, \rhv, \rho^{\delta\texttt{v}}, \forall i \in \I$
\State Initialize $x_i, \ep, \eq, \LocalPriceP, \LocalPriceQ$, $\forall i \in \I$; set $k \gets 1$
\While{not converged}
    \ForAll{agents $i \in \I$ \textbf{in parallel}}
        \State Calculate $\LocalPriceP[k], \LocalPriceQ[k]$ using Eqs.~\eqref{eq:Plambda_update}--\eqref{eq:Qlambda_update}.
        \State Update $x_i[k]$ by solving problem \eqref{LocalProblem}.
        \State Calculate $\ep[k], \eq[k]$ using Eqs.~\eqref{eq:ResidualP}--\eqref{eq:ResidualQ}.
        \State Send $v_{i,t}[k], \ep[k], \eq[k], \LocalPriceP[k], \LocalPriceQ[k]$ to all neighbors $j \in \Ni$.
    \EndFor
    \State $k \gets k+1$
\EndWhile
\State \Return $x_i, \LocalPriceP, \LocalPriceQ, \ep, \eq$, $\forall i \in \I$
\end{algorithmic}
\end{algorithm}

The decision variables of the problem are
\begin{equation}
    \mathcal{X}_i = \left\{x_i, \ep, \eq, \UpSlackVoltage, \DownSlackVoltage, \Vijt \right\},
\end{equation}
whereas the necessary parameters $\LocalPriceP, \LocalPriceQ$, decisions of the previous iteration $\p[i][k-1],\q[i][k-1], v_{j,t}[k-1]$ and neighbors residuals $\jresidualP[k-1], \jresidualQ[k-1] \forall j \in \Ni$ are updated before the problem is solved based on the exchanged communication signals with $i$'s neighbors, as defined in Algorithm \ref{alg:ConsensusAlgorithm}.
The algorithm is deemed converged when
\begin{align*}
  \big\|\LocalPriceP[k] - \lambda^{\texttt{p}}_{j,t}[k-1]\big\|_2
    &\leq \varepsilon_\lambda, &&\forall i \in \I,\ j \in \Ni,
    \\
  \big\|\LocalPriceQ[k] - \lambda^{\texttt{q}}_{j,t}[k-1]\big\|_2
    &\leq \varepsilon_\lambda, &&\forall i \in \I,\ j \in \Ni,
    \\
  \big\|\ep[k]\big\|_2,\ \big\|\eq[k]\big\|_2
    &\leq \varepsilon_\delta, &&\forall i \in \I.
\end{align*}
Observe that Algorithm \ref{alg:ConsensusAlgorithm} can readily be used for solving the voltage-agnostic plain economic dispatch \ED\ benchmark problem as well, by simply setting the weights $\rhv$ and $\rho^{\delta\texttt{v}}$ of Eq. \eqref{eq:F_indiv_voltage} equal to zero.
%select what Algorithm~\ref{alg:ConsensusAlgorithm} solves. Setting both to zero makes $\mathcal{V}_i \equiv 0$ and decouples the voltage variables from $x_i$, so that \eqref{LocalProblem} reduces to the minimization of $\mathcal{L}_i$: this is the voltage-agnostic variant, whose fixed points are optimal for \ED\ (Theorem~\ref{theor:optimality}). Any $\rhv > 0$ activates the voltage penalty, while a large $\rho^{\delta\texttt{v}}$ drives the agents' voltage estimates into agreement; the fixed points are then optimal for the penalized problem, and Theorem~\ref{theor:error_quant} bounds their economic suboptimality considering the surrogate's accuracy. Both variants are benchmarked in Section~\ref{Sec:Results}.}
%
We conclude this subsection by establishing the theoretical properties of
Algorithm~\ref{alg:ConsensusAlgorithm}. 
We first need to state certain assumptions:

\begin{assumption}[Communication graph]\label{ass:graph}
The graph $\graph = (\I,\E)$ is connected and contains a self-loop at every node. The communication matrix $\mathbf{W}$ is nonnegative and doubly stochastic, i.e.\ $\mathbf{W}\mathbf{1} = \mathbf{1}$ and $\mathbf{1}^\top \mathbf{W} = \mathbf{1}^\top$, with $W_{i,j} > 0$ if and only if $(i,j) \in \E$.
\end{assumption}

\begin{assumption}[Regularity]\label{ass:convex}
For every $i \in \I$, the local cost $F_i(x_i) \coloneqq \sum_{t \in \T}f_i(\gen)$ is convex and continuously
differentiable, and the local feasible set $\Omega_i \coloneqq \{x_i : \eqref{eq:gen_lims}\text{--}%
\eqref{eq:agent_reactive_power_balance} \text{ hold}\}$ is nonempty, closed
and convex.
\end{assumption}

\begin{assumption}[Initialization]\label{ass:init}
The residual estimates are initialized at the agents' net injections, i.e.\
$\ep[0] = \p[i][0]$ and $\eq[0] = \q[i][0]$ for all $i \in \I$, $t \in \T$.
\end{assumption}
\vspace{4pt}

%Finally, let \ED\ denote the plain \emph{Economic Dispatch problem}, i.e.\ \eqref{centralized} stripped of the voltage constraints \eqref{eq:voltage_equation} and \eqref{eq:voltage_limits}: minimize $\sum_{t \in \T} \sum_{i \in \I} f_i(\gen)$ over $x_i \in \Omega_i$, $\forall i \in \I$, subject to \eqref{eq:active_power_balance}\text{--}\eqref{eq:reactive_power_balance}.

\noindent We first turn to the benchmark problem (ED) and the voltage-agnostic version of Algorithm \ref{alg:ConsensusAlgorithm}, before generalizing to problem \eqref{centralized}. The following theorem establishes the Algorithm's optimality for problem (ED).

\begin{theorem}[Optimality of fixed points]\label{theor:optimality}
Let Assumptions~\ref{ass:graph}--\ref{ass:init} hold and $(x^\star, \bm{\lambda}^\star, \bm{\delta}^\star)$ be a fixed point of Algorithm~\ref{alg:ConsensusAlgorithm}, with $\rhv,\rho^{\delta\texttt{v}} = 0$. Then %the agents' multiplier proxies reach a common value per time step, i.e.\ 
there exist $\bar{\lambda}^{\texttt{p}}_t, \bar{\lambda}^{\texttt{q}}_t \in \mathbb{R}$ such that $\lambda^{\texttt{p}\star}_{i,t} = \bar{\lambda}^{\texttt{p}}_t$ and $\lambda^{\texttt{q}\star}_{i,t} = \bar{\lambda}^{\texttt{q}}_t$ for all $i \in \I$, $t \in \T$. Moreover, $x^\star$ is an optimal solution of problem \ED, and $(\bar{\lambda}^{\texttt{p}}_t, \bar{\lambda}^{\texttt{q}}_t)_{t \in \T}$ are optimal Lagrange multipliers of \eqref{eq:active_power_balance}--\eqref{eq:reactive_power_balance}, with zero duality gap.
\end{theorem}

\begin{IEEEproof}
The proof is given in Appendix~\ref{Sec:Apndx_A}.
\end{IEEEproof}

\vspace{4pt}
%For the case where the voltage penalty is carried inside $F_i$, i.e.\ the local problem \eqref{LocalProblem} as actually solved by Algorithm~\ref{alg:ConsensusAlgorithm}, we make the following Remark.

%Theorem~\ref{theor:optimality} applies to problem \ED. It extends to \eqref{LocalProblem} as actually solved, with the voltage variables appended to $x_i$ and the convex penalty \eqref{eq:F_indiv_voltage} added to $F_i$, provided the agents agree on their shared voltage estimates at the fixed point, i.e.\ $\Vijt = v^{(j)}_{j,t}$ for all $j \in \Ni$. The fixed point is then optimal for the joint economic and voltage-related cost. Considering the economic cost we present the following. 
%This agreement is enforced only approximately, by the penalty of weight $\rho^{\delta\texttt{v}}$, and strongly holds at the limit $\rho^{\delta\texttt{v}} \to \infty$.

We now turn to problem \eqref{centralized} and, correspondingly, to the full version of Algorithm \ref{alg:ConsensusAlgorithm} where $\rhv,\rho^{\delta\texttt{v}} > 0$ and the voltage variables are appended to $x_i$. We are now in position to establish the optimality guarantee of Algorithm \eqref{alg:ConsensusAlgorithm} as a function of the error of the voltage surrogate model $\phi(\cdot)$ of Eq. \eqref{eq:abstract_voltage_estimator}.
We first state two further assumptions formally.
%Naturally, the penalty term $\mathcal{V}_i$ shifts the optimum to $\xa$ if voltage estimates sit outside the admissible range, trading economic cost against voltage deviations and neighbors' voltage estimates. Below are some further assumptions necessary towards analyzing the impact of the voltage surrogate impact.
\begin{assumption}[Voltage opinion consensus]\label{ass:volt_cons}
At every fixed point of Algorithm~\ref{alg:ConsensusAlgorithm} the agents agree on their shared voltage estimates, $\Vijt = v^{(j)}_{j,t}$ for all $i \in \I$, $j \in \Ni$, $t \in \T$.
\end{assumption}

\vspace{3pt}
\noindent Assumption~\ref{ass:volt_cons} is enforced only approximately, by the penalty of weight $\rho^{\delta\texttt{v}}$, and holds exactly in the limit $\rho^{\delta\texttt{v}} \to \infty$. To quantify the impact of the voltage surrogate model on the economic cost alone, we additionally assume:
\begin{assumption}[Voltage surrogate accuracy]\label{ass:sur_acc}
There exists $\epsv \ge 0$ such that $|\phi_i - \Phi_i| \le \epsv$, with $\Phi_i$ and $\phi_i$ as in \eqref{eq:voltage_equation} and \eqref{eq:abstract_voltage_estimator}, for all $i \in \I$, $t \in \T$ and all injections feasible for problem \ED.
\end{assumption}

\vspace{3pt}
\noindent Our main theoretical result follows.
%We are now able to connect the error introduced by the surrogate model to the overall algorithm's optimality:
\begin{theorem}\label{theor:error_quant}
Let Assumptions~\ref{ass:graph}--\ref{ass:sur_acc} hold, let $\xa$ be a fixed point of Algorithm~\ref{alg:ConsensusAlgorithm}, with  $\rhv,\rho^{\delta\texttt{v}} > 0$, and corresponding system cost $C^\star = \sum_{i \in \I} F_i(\xa_i)$,
and let \eqref{centralized} be feasible with optimal value $H^\star(\xb)$. Then
\begin{equation}
  C^\star \;\le\; H^\star + 2 \rhv \, |\T| \, \Gamma \, \epsv^2 ,
  \label{eq:corollary_bound}
\end{equation}
where $\Gamma := \sum_{i \in \I} |\Ni|$ is the number of replicated voltage terms.
\end{theorem}

\begin{IEEEproof}
The proof is given in Appendix~\ref{Sec:Apndx_B}.
\end{IEEEproof}

\subsection{Learning a Surrogate Voltage Mapping} \label{sec:surrogate}

In this subsection, we instantiate the voltage estimator of Eq.~\eqref{eq:abstract_voltage_estimator}, which is an essential element of Algorithm~\ref{alg:ConsensusAlgorithm}.
It is important to note that the estimator $\phi(\cdot)$ enters each agent's optimization problem \eqref{LocalProblem} through constraint \eqref{eq:voltage_individual_estimation}, and it is therefore necessary that $\phi(\cdot)$ remains linear in the decision variables $x_i$.
Under these considerations, we use a data-driven Sparse Group LASSO (SGL) regression \cite{Simon01012013}, a successor of the group LASSO \cite{LASSO_grouped/j.1467-9868.2005.00532.x} method, to extract information about the substrate electrical network in a data-driven manner. 
Specifically, we approximate the unknown mapping of Eq. \eqref{eq:voltage_equation} through a surrogate linear model
\begin{align}
    \ve=&  \sum_{j \in \I} \left( \bp \ p_{j,t} + \bq q_{j,t}\right) + \bz, &&\forall i \in \I,  \label{eq:voltage_estimation}
\end{align}
The intercept parameter is $\bz$, while $\bp$ and $\bq$ are the active and reactive power coefficients capturing the effect of $j$'s power injections on $i$'s voltage.

%Through this process, a best-fit communication graph is imposed on top of the electrical network. The best-fit communication graph defines, for each node, the best linear approximation, according to SGL cost, of its voltage as a function of the active and reactive power at that node and a selected set of neighboring nodes, enabling the voltage to be enforced as a constraint within an admissible range. This regularization technique is adequate in the sense that it can promote sparsity in the coefficient matrices while grouping their elements in subsets which may represent bidirectional communication channels between nodes. The resulting coefficients indicate a communication graph of the agents, and the linear voltage approximation will be used as a constraint in the presented Consensus-Based algorithm, which is then upgraded to a data-based voltage-aware consensus EDP.

%As already explained, we estimate the logarithm of voltage  ~\cite{Zhigang_Log} in a linear fashion to improve accuracy over a simple linear estimation of the voltage, while maintaining a function that is adequate for our optimization problem and maintain its convex nature. This relation comes from the $\phi_i$ mapping, introduced in \eqref{eq:voltage_equation}, that we consider for this study. Since we bound the voltage both above and below, and the logarithm is strictly increasing, the limits transform directly into bounds on the log-voltage, introducing no nonlinearity.

%Let us suppose that historical data that cover sufficient spectrum of the grid's capabilities is obtained. 
The model is trained using available historical measurements of nodal power injections and the corresponding voltages, collected during normal system operation. Following \cite{Zhigang_Log}, we use the transformed voltage $\tilde v = 1 + \ln v$ measurements, which keeps the estimator linear in the decision variables and the admissible range centered near unity; since the logarithm is strictly increasing, the limits \eqref{eq:voltage_limits} map directly onto bounds on $\tilde v$. 
Let $\pairs = \{(i,j) \in \mathcal{I}^2 : i < j\}$ be the set of bus pairs, and let $m \in \mathcal{M}$ denote a data point $\left( \left(\labelP \right)_{j \in \I}, \left(\labelQ \right)_{j \in \I}, \labelV\right)$, where $\labelV$ is the transformed voltage at node $i$. The set $\mathcal{N}$ of such data points is used to fit the parameters $y \coloneqq \Bigl( \bigl(\bp,\, \bq \bigr)_{(i,j) \in \pairs},\; \bigl(\bz\bigr)_{i \in \I} \Bigr)$ by solving
\begin{equation}
\min_{y} \Big \{
  %\frac{1}{|\mathcal{M}||\mathcal{I}|} 
  J(y) + \nu_1 \Bone + \nu_2 \Btwo \Big \},
  \label{eq:sgl}
\end{equation}
where the main term in the objective function is
\begin{multline}
    J(y) = \frac{1}{|\mathcal{M}||\mathcal{I}|} 
    \sum_{m\in \mathcal{M}} \sum_{i\in \I}
  \Bigg( \labelV - \\ \Bigg(\sum_{j \in \I} \left( \bp \ p^{(m)}_j + \bq q^{(m)}_{j}\right) + \bz \Bigg) \Bigg)^2
\end{multline}
and there are two more penalization terms, defined as follows. First,
\begin{equation}
  \Bone = \sum_{i \in \I} \Big( \sum_{j \in \I}
  \left( |\bp| + |\bq| \right) + |\bz| \Big)
\end{equation}
is the $L_1$ penalty, which promotes sparsity of individual coefficients. 
For the last term, let
$\Bg = [\bp,\, \bq,\, \bp[j,i],\, \bq[j,i]]^\top$ the coefficient group of
pair $(i,j)$. Then
\begin{equation}
  \Btwo = \sum_{(i,j) \in \pairs} w_{i,j}
  \sqrt{(\bp)^2 + (\bq)^2 + (\bp[j,i])^2 + (\bq[j,i])^2}
\end{equation}
is the group ($L_{2,1}$) penalty, which is a $w_{i,j} > 0$ weighted sum of $\|\Bg\|_2$ and promotes sparsity at the level of bus pairs.
To handle the non-differentiable group norm, we introduce an auxiliary variable $\gamma_{i,j}$ for each pair $(i,j) \in \pairs$ and apply an epigraph reformulation: replacing $\|\Bg\|_2$ by $\gamma_{i,j}$ in the objective and imposing the second-order cone constraint $\|\Bg\|_2 \leq \gamma_{i,j}$ yields the equivalent conic problem

\begin{align}
\min_{y'} &\left\{ J(y')  + \nu_1 \Bone
    + \nu_2 \sum_{(i,j) \in \pairs} w_{i,j}\, \gamma_{i,j} \right\}
    \label{eq:group_lasso} \\
\text{s.t.} &\quad
  \|\Bg\|_2 \leq \gamma_{i,j}, \quad \forall (i,j) \in \pairs,
    \label{eq:soc}
\end{align}
where $y' \coloneqq \bigl( y,\, (\gamma_{i,j})_{(i,j) \in \pairs} \bigr)$. 

%This process creates two main effects apart from its main purpose, which is the minimization of the mean squared error between the data set and the estimator. Firstly, it penalizes the number of interactions between the variables, and secondly, it provides the possibility to group the variables into meaningful subsets for the presented case. In this way, we can penalize the coefficients in a grouped manner so that a channel between two agents is activated or not in a bidirectional manner.

%\textcolor{red}{Description of decomposition of problem defined by (39)--(40). Do we need such detailed explanation?}
Although the fitting problem \eqref{eq:group_lasso}-\eqref{eq:soc} is a centralized optimization problem, it only needs to be solved offline.
%(once, or with subsequent re-fitting performed periodically). 
Moreover, if a decentralized solution is required also for this offline process, the structure of~\eqref{eq:group_lasso} is well suited to decomposition~\cite{conejo2006decomposition} or federated learning settings~\cite{mcmahan2017communication}, where federated composite optimization methods~\cite{yuan2021federated} can handle nonsmooth penalties.
%Moreover, if a decentralized solution is required also for this offline process, the structure of~\eqref{eq:group_lasso} is well suited to decomposition~\cite{conejo2006decomposition}. If each bus $i$ decides upon its own coefficients $\bp$, $\bq$, $j \in \I$, and its intercept $\bz$, both the data-fitting term and the $L_1$ penalty are separable across buses. However, a complicating term, $\gamma_{i,j}$, appears in the cost function~\eqref{eq:group_lasso} and the cone constraint~\eqref{eq:soc}. Introducing, at each bus $i$, local copies of $\bp[j,i]$, $\bq[j,i]$ and $\gamma_{i,j}$, and enforcing their equality with the originals, turns this coupling into a set of linear consensus constraints. These can be relaxed through an augmented Lagrangian, leading to a distributed ADMM scheme~\cite{boyd2011distributed, molzahn2017survey} in which each bus solves a small local second-order cone program and exchanges only these copies with the buses it is paired with. The same structure also fits a federated learning setting~\cite{mcmahan2017communication}, where federated composite optimization methods~\cite{yuan2021federated} can handle nonsmooth penalties. By using such decentralized training schemes, the entire framework (from surrogate modeling to operation) can be fully decentralized. 
%Their implementation is, however, outside the scope of this work.

%\vspace{50pt}

\subsection{Constructing the Communication Graph and Communication Matrix}\label{sec:graph}

In this subsection, we instantiate the communication graph $\graph$ (equivalently, $\Ni$ for all $i \in \I$) required by Algorithm~\ref{alg:ConsensusAlgorithm}, deriving it from the coefficients of the surrogate $\phi(\cdot)$ fitted in Subsection~\ref{sec:surrogate}, and conclude with the construction of the communication matrix $\mathbf{W}$.
%Our contribution here lies in the fact that we couple the communication matrix W of Algorithm~\ref{alg:ConsensusAlgorithm} with the voltage estimation mapping, $\phi(\cdot)$, to render the communication process among agents efficient. 

The objective is to retain a sufficient set of edges while eliminating unnecessary ones, where an edge $(i,j)$ corresponds to the coefficient group $\Bg$ defined in Subsection~\ref{sec:surrogate}. In principle, the group penalty in \eqref{eq:group_lasso} already performs this selection, since an edge should be retained if and only if $\Bg \neq \mathbf{0}$. In practice, however, we cannot use the solution of problem \eqref{eq:group_lasso} directly for two reasons. First, since the problem is solved numerically, coefficients that should vanish typically take small but nonzero values. Second, some nodes have all their coefficients close to zero for physical reasons. For example, the upstream grid imposes the slack-node voltage, so it does not respond to nodal injections. Consequently, a nonzero test would retain almost all pairs, while a simple threshold on the coefficients could isolate such nodes and render the graph disconnected. We therefore select the edges through Algorithm~\ref{alg:graph_construction}.

%where $g\in \mathbb G$ denotes a group of two agents $(i,j)$ and $\mathbb G$ is the set of all such groups. Both terms contribute to the cost function of the SGL problem. However, 
%The cost function of problem (\ref{eq:group_lasso}) represents a SGL regression, which is used to find the optimal values of the aforementioned coefficients, as it can create an effect of activating or suppressing a chosen group of parameters of $\mathrm{b}_{p}, \mathrm{b}_{q}$ entries. Thus, we choose to group these parameters as node pairs. If we suppose possible neighboring nodes $i,j$ we group $\mathrm{b}_{ij}= \{\bp, \bq\}$ and $\mathrm{b}_{ji} = \{\bp[j,i], \bq[j,i]\}$ to establish, if beneficial for the voltage estimation of both nodes, a bidirectional communication channel which will take into account all neighbors' variables. By solving this problem, we can construct a communication graph without any information about the grid's topology or parameters. 

% To improve the performance of the linear estimator, we do the following change of variables $\tilde{V} = 1+lnV$ \cite{Zhigang_Log}, which does not affect the convex nature of the optimization problem. Limits of the constraint (\ref{eq:voltage_limits_compact}) are adapted similarly. Other ways of estimating voltage surrogate constraints can be applied in our formulation as long as they maintain the convex nature of the problem.

First, let us introduce the connection importance matrix $\mathbf{C}$, which is defined as a positive symmetric matrix that quantifies the relevance of potential interactions between nodes and is a necessary input for Algorithm~\ref{alg:graph_construction}. We define: 
\begin{align}
\tilde{C}_{ij} = \max_{m \in \mathcal{M}}\left ( | \bp \ p^{(m)}_{j} + \bq q^{(m)}_{j}| \right)
\end{align}

\noindent and enforce:
\begin{align}
C_{ij} = \max(\tilde{C}_{ij}, \tilde{C}_{ji}) \label{eq:importance_metric}
\end{align}
This preserves the symmetric nature of the edges.

\begin{algorithm}[t]
\caption{Construction of the Communication Graph}
\label{alg:graph_construction}
\begin{algorithmic}[1]
\Require Importance matrix $\mathbf{C}$, threshold $\vartheta$, min degree $d_{\min}$
\Ensure Connected communication graph $\graph = (\I, \E)$
\State $\E \gets \{(i,i) : i \in \I\}$ \Comment{self-loops}
\For{each node $i \in \I$} \Comment{Stage 1}
    \State Sort $j \in \I \setminus \{i\}$ so that $C_{i,j_1} \leq \dots \leq C_{i,j_{|\I|-1}}$
    \State $\kappa^\star \gets \min\big\{ \kappa : \sum_{\ell=1}^{\kappa} C_{i,j_\ell} > \vartheta \big\}$
    \State $\E \gets \E \cup \{(i,j_\ell), (j_\ell,i) : \ell \geq \kappa^\star\}$
\EndFor
\While{$\graph$ is not connected} \Comment{Stage 2}
    \State $(i^\star, j^\star) \gets$ pair with the largest $C_{i,j}$ such that $i$ and $j$ are in different components
    \State $\E \gets \E \cup \{(i^\star,j^\star), (j^\star,i^\star)\}$
\EndWhile
\For{each node $i \in \I$} \Comment{Stage 3}
    \While{$|\Ni| - 1 < d_{\min}$}
        \State $j^\star \gets \arg\max \{ C_{i,j} : j \notin \Ni \}$
        \State $\E \gets \E \cup \{(i,j^\star), (j^\star,i)\}$
    \EndWhile
\EndFor

\State \Return $\graph \text{ and } \bp, \bq, \bp[j,i], \bq[j,i] \forall (i,j) \in \E$
\end{algorithmic}
\end{algorithm}

The presented Algorithm~\ref{alg:graph_construction} comprises three stages: In the first stage, cumulative importance is evaluated with respect to the indicated metric in \eqref{eq:importance_metric}, and the $ |\I|- (\kappa^\star+1)$ most relevant interactions are retained as edges in the communication graph. A threshold $\vartheta$ controls the acceptable accuracy of the voltage estimation.
In the second stage, if $ \graph$ is not connected, the most valuable connections between components are established. In the third stage, nodes with degree below the prescribed minimum increase their connectivity according to the connection importance matrix. The retained interactions that result in the mapping $\phi$ are considered bidirectional communication links in the resulting graph $\graph$. 
%We later use the obtained graph $\mathcal G$ for our Consensus-Based approach.

%\textcolor{blue}{Algorithm~\ref{alg:ConsensusAlgorithm} requires} a weight matrix $\mathbf{W} = [W_{i,j}]$ that is nonnegative, doubly stochastic, and consistent with the graph, i.e., $W_{i,j} > 0$ if and only if $(i,j) \in \E$. Both stochasticity properties are needed. Row stochasticity ($\mathbf{W}\mathbf{1} = \mathbf{1}$) ensures that averaging preserves agreement among the dual variables $\LocalPriceP, \LocalPriceQ$, while column stochasticity ($\mathbf{1}^\top \mathbf{W} = \mathbf{1}^\top$) ensures that the local residual estimates $\ep, \eq$ always sum to the true community imbalance (Lemma~\ref{lem:conserv} in Appendix~\ref{Sec:Apndx_A}). Together, they guarantee that the algorithm's fixed points satisfy the community power balance with a common dual-variable value across all agents (Lemma~\ref{lem:lambda}). 

%Related consensus schemes~\cite{Papadaskalopoulos_2021, Wu_2026} assume such a matrix but do not address how to obtain it for a general graph.

To construct the doubly stochastic matrix $\mathbf{W}$ that is required by Algorithm~\ref{alg:ConsensusAlgorithm}, we propose these two steps: we first form a row-stochastic matrix from the graph's Laplacian, and then apply the Sinkhorn--Knopp algorithm~\cite{sinkhorn1967concerning}, which alternately normalizes rows and columns until the matrix is doubly stochastic. Both steps require only information exchanged between neighbors, so the construction preserves the scheme's decentralized nature. Other constructions with the same properties, such as Metropolis or maximum-degree weights~\cite{xiao2005scheme} or weights optimized for convergence speed~\cite{xiao2004fast}, could be used instead. We detail the construction we adopt in Appendix~\ref{Sec:Apndx_C}.

This methodology for obtaining $\graph$ and $\mathbf{W}$, together with the one for obtaining $\phi_i(\cdot)$, completes the full instantiation of the proposed consensus-based coordination scheme, as given by Algorithm \ref{alg:ConsensusAlgorithm}. In the next section, we empirically evaluate the proposed scheme's properties.

\section{Numerical Evaluation} \label{Sec:Results}
In this section, we first describe the setup and the scenarios used to evaluate the proposed approach in Subsection~\ref{subsec:UseCase}. Subsection~\ref{subsec:LassoNum} reports the SGL results, and Subsection~\ref{subsec:Cons-Numerical-Ev} evaluates the proposed VCED scheme. Finally, Subsection~\ref{subsec:SensitivityAnalysis} concludes with a sensitivity analysis for a high-DER scenario and the voltage penalty parameter $\rho^\texttt{v}$.

\subsection{Use Cases and Experiment Setup} \label{subsec:UseCase}

The network used to test the proposed methodology is an enhanced version of the standard IEEE-33 medium-voltage network, as presented in \cite{Hatziargyriou_2021}, including cost functions and resource capacities. All loads are treated as flexible except at buses 11 and 22, whose demand follows a fixed profile. The cost coefficients are divided by a factor of ten so that the linear term is of unit order.
%and the constant terms $c_{0,i}$ are excluded from the normalization of the relative metrics: at roughly $95\%$ of the total cost and invariant to the schedule, they would understate every reported gap by more than an order of magnitude.
%This improves the numerical conditioning of the local problems and makes the reported multipliers directly interpretable as prices. 
The generators at the feeder ends have active power capacity only; reactive power is supplied by the generator at bus 0 (the slack node), whose capacity covers the maximum load, and by the two reactive power compensators.
\begin{comment}
\begin{figure}[t]
    \centering
    \includegraphics[width=0.4\textwidth]{images/ieee33_enhanced.png}
    \caption{IEEE-33 bus medium voltage network}
    \label{fig:IEEE-33_enhanced}
\end{figure}
\end{comment}
Our experiments are organized as follows:
\subsubsection{Surrogate voltage model with SGL}
The setup for testing our surrogate voltage model via SGL comprises three steps: a) data point generation, b) power flow runs for all obtained data points, and c) feasibility filtering, which together complete the data set generation.

\begin{comment}
    All agents, except the slack node representing the grid connection, are modeled as PQ nodes. Demand levels are sampled by randomly selecting values for each load within its operational limits, and reactive power is determined based on a constant power factor assumption, as reported in \cite{Hatziargyriou_2021}. Generators and energy storage units are sampled similarly. The per-resource samples are then combined into joint instances, each specifying $p$ and $q$ at every node, while the slack node is treated as a $V\theta$ node. We perform power flow calculations for all instances and only keep feasible solutions until we collect a data set of $N = 2000$ points. This is then split into a training set (80\%) and a held-out test set (20\%).%; all reported errors are computed on the test set.
\end{comment}

All agents are modeled as PQ nodes, except the slack node modeled as a $V\theta$. Demand levels are randomly sampled within operational limits, and reactive power follows the assumption of constant power factor, as reported in \cite{Hatziargyriou_2021}. Generators and energy storage units are sampled similarly. We then perform power flow calculations for all instances and only keep feasible solutions until we collect a data set of $N$ points. This is split into a training set (80\%) and a held-out test set (20\%).%; all reported errors are computed on the test set.

The data set is generated by simulation because no measurement record exists for a benchmark network. In a deployment, the same quantities (each agent's own injections and the voltage at its own bus) are available directly from metering. The power flow here plays the role of the physical system that produces those measurements: neither the fitted model nor Algorithm~\ref{alg:ConsensusAlgorithm} has access to it, and the admittance matrix enters nowhere in the scheme.

%The resulting data set spans the operating region of all resources and is therefore well suited to identifying the voltage--power sensitivities. Measurements collected in operation would be more strongly correlated across nodes, which is known to make group selection harder; assessing the estimator under such conditions is left for future work.

\subsubsection{VCED with low DERs}
The benchmark scenario for this evaluation follows the resource configuration presented in \cite{Hatziargyriou_2021}. Two of the four generators at the feeder ends are scheduled generators, representing forecasted PV units with a rated capacity of \SI{0.2}{\mega\watt} each, operating between $89\%$ to $98\%$ of their rated capacity. All dispatchable loads are constrained to their average reported consumption over the time horizon, with a maximum capacity of $125\%$ of the peak consumption also reported in~\cite{Hatziargyriou_2021}. The two reactive power compensators, included as reactive power generators, respond only to the Lagrange multiplier cost.
    
\subsubsection{VCED with high DERs and sensitivity w.r.t. $\rhv$}
The final group of experiments comprises two parts: a) a high-DER scenario, which tests the scheme under a resource generation substantially elevated from the one the surrogate was trained on, and b) a sensitivity analysis with respect to $\rhv$ in the benchmark scenario.The high-DER scenario raises installed PV from \SI{0.4}{\mega\watt} to \SI{4.2}{\mega\watt} --- four \SI{1}{\mega\watt} units at buses without scheduled generation, plus one retained \SI{0.2}{\mega\watt} benchmark unit --- and adds two energy storage units. 
%The resulting configurations are presented in Table~\ref{tab:Comparison}. %This alters the nature of the problem, not just its magnitude. 
%While the benchmark is dominated by undervoltage at the ends of the feeders, the increased injection reverses power flow during high-generation hours, and the binding constraint becomes the upper limit of $1.05$~p.u. 
Finally, we analyze the impact of $\rhv$ under the low-DER scenario. 

The scenarios above are constructed to stress the scheme rather than to represent typical operation: generation is held close to rated output over the whole horizon, the entire load is flexible, and the high-DER case reaches a PV capacity comparable to the peak load. Representative daily profiles would place considerably less pressure on the voltage limits.

\begin{comment}
\begin{table}[t]
    \centering
    \caption{Low-DER and High-DER Scenarios characteristics. 
    %Load requirement is identical in both. PV fleet is expanded and two energy storage units are added. Available storage energy is the range between state-of-charge limits.
    }
    \label{tab:Comparison}
    \begin{tabular}{lcc}
        \toprule
        Quantity & Low-DER & High-DER \\
        \midrule
        Scheduled PV units                     & 2      & 5           \\
        Total PV installed capacity [MW]       & 0.40   & 4.20        \\
        Average PV output [MW]                 & 0.38   & 3.97        \\
        \midrule
        Average load [MW]                      & 2.96   & 2.96        \\
        Peak load [MW]                         & 4.46   & 4.46        \\
        \midrule
        Energy storage units                   & 0      & 2           \\
        Storage power, charge / discharge [MW] & ---    & 0.36 / 0.34 \\
        Storage energy capacity [MWh]          & ---    & 1.03        \\
        Available storage energy [MWh]            & ---    & 0.75        \\
        \midrule
        Average PV / average load              & 12.8\% & 134.1\%     \\
        PV capacity / peak load                & 9.0\%  & 94.1\%      \\
        \bottomrule
    \end{tabular}
\end{table}
\end{comment}

Table~\ref{tab:params} lists the selected values for all parameters used to perform the experiments. All simulations were carried out in Python 3.9.18 using Pyomo 6.7.0 with Gurobi 10.0.3, which solves both the offline SGL fit and the per-agent second-order cone problems, on a laptop with an Intel Core i5-8365U CPU @ \SI{1.60}{\giga\hertz} and \SI{16}{\giga\byte} of RAM, running Windows 11. 
%Finally, since the agents' problems are independent within an iteration, all results that indicate computational time, assume parallel solver time (elapsed time), computed as the sum over iterations of the maximum per-agent solve time, i.e.\ the time a fully parallel implementation would require; communication latency is not included.

\begin{table}[t]
  \centering
  \caption{Parameter Settings}
  \label{tab:params}
  \begin{tabular}{llc}
    \toprule
    Symbol & Description & Value \\
    \midrule
    $|\mathcal{M}|$              & Data points (train / test) & 2000 (1600 / 400) \\
    $\vartheta$ & Importance threshold, Alg.~\ref{alg:graph_construction} & $5\cdot10^{-3}$\\
    $d_{\min}$ & Minimum node degree & 2\\
    $|\T|$ & Horizon length (1-h steps) & 8 \\
    $[\underline{V},\overline{V}]$ & Admissible voltage range [p.u.] & [0.95, 1.05] \\
    $\rhop, \rhoq$ & Residual penalties & 2, 2 \\
    $\rhv$ & Voltage-deviation penalty & 4 \\
    $\rho^{\delta\texttt{v}}$ & Voltage-consensus penalty & 1 \\
    $\epsp, \epsq$ & Dual step sizes & $5\cdot10^{-2}$ \\
    $\varepsilon_\lambda$ & Multiplier agreement tolerance & $10^{-4}$ \\
    %$\varepsilon_\delta$ & Residual tolerance [MW] & $10^{-3}$ \\
    $\Delta^{\mathrm{gap}}$ & Acceptable relative duality gap & $10^{-1}\,\%$ \\
    $\varepsilon_{\mathrm{SK}}$ & Sinkhorn--Knopp tolerance & $10^{-12}$ \\
    $k_{\max}$ & Iteration budget & 2000 \\
    \bottomrule
  \end{tabular}
\end{table}

\subsection{Empirical Evaluation of the Surrogate Voltage Mapping}\label{subsec:LassoNum}

\begin{figure}[t]
    \centering
    
    \begin{subfigure}{0.24\textwidth}
        \centering
        \includegraphics[width=\linewidth]{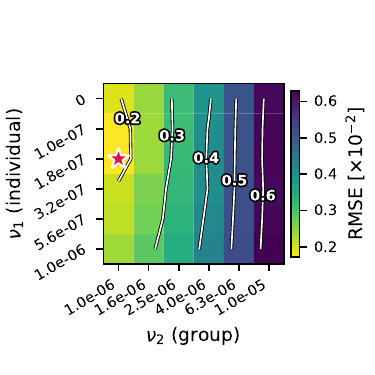}
        \caption{}
    \end{subfigure}
    \hfill
    \begin{subfigure}{0.24\textwidth}
        \centering
        \includegraphics[width=\linewidth]{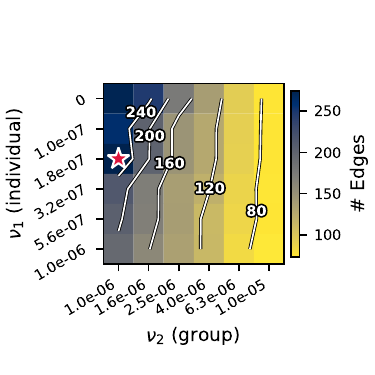}
        \caption{}
    \end{subfigure}
    \caption{Root Mean Squared Error (a) and Number of Edges in Communication Graph (b) for $\nu_1 \in \{0\}\cup  [ 10^{-7}, 10^{-6}]$ and $\nu_2 \in [ 10^{-6}, 10^{-5}].$ The star marks the RMSE optimum.}
    \label{fig:Errors}
\end{figure}

In this subsection, we evaluate the surrogate voltage mapping in isolation.
By solving Problem \eqref{eq:group_lasso} on the data points in Table~\ref{tab:params} and varying the penalization parameters $\nu_1$ and $\nu_2$, we report the results in Fig.~\ref{fig:Errors}. These include the RMSE and the number of edges in each communication graph, shown in a 2-D color diagram with $(\nu_1, \nu_2)$ combinations.

%The proposed parameter selection strategy should not minimize RMSE; instead, a maximum error tolerance is defined, and parameters are selected to minimize connectivity while remaining within this bound. This approach is conservative, as it reflects a robust logic of the voltage estimation and control framework.
As shown in Fig.~\ref{fig:Errors}, the estimation error increases with $\nu_1$ and $\nu_2$. Also, the number of communication edges decreases, so the two penalties trade estimation accuracy against the connectivity of the induced graph. In deployment, we choose the operating point along this trade-off by fixing an acceptable estimation error and taking the sparsest graph that meets it. In the evaluation that follows, we fix
$\nu_1 = 1.78\cdot10^{-7}$, the value minimizing the empirical RMSE, and vary $\nu_2$ to span a range of graph densities, which lets us directly examine the effect of sparsity on the dispatch and meeting of voltage constraints.
\begin{comment}
\begin{figure}[t]
    \centering
    
    \begin{subfigure}{0.11\textwidth}
        \centering
        \includegraphics[width=\linewidth]{images/LASSO/Communication_graphs/Substrate_Electrical_Network.pdf}
        \caption{}
    \end{subfigure}
    \hfill
    \begin{subfigure}{0.11\textwidth}
        \centering
        \includegraphics[width=\linewidth]{images/LASSO/Communication_graphs/comgraph_nu1_1.78e-07_nu2_1.00e-06.pdf}
        \caption{}
    \end{subfigure}
    \hfill
    \begin{subfigure}{0.11\textwidth}
        \centering
        \includegraphics[width=\linewidth]{images/LASSO/Communication_graphs/comgraph_nu1_1.78e-07_nu2_2.51e-06.pdf}
        \caption{}
    \end{subfigure}
    \hfill
    \begin{subfigure}{0.11\textwidth}
        \centering
        \includegraphics[width=\linewidth]{images/LASSO/Communication_graphs/comgraph_nu1_1.78e-07_nu2_1.00e-05.pdf}
        \caption{}
    \end{subfigure}

    \caption{(a) Substrate electrical graph, and communication graphs obtained with $\nu_1 = 1.78\cdot10^{-7}$ and (b) $\nu_2 = 1\cdot10^{-6}$, (c) $\nu_2 = 2.51\cdot10^{-6}$, (d) $\nu_2 = 1\cdot10^{-5}$, corresponding to $275$, $148$ and $75$ edges respectively.}
    \label{fig:electr_and_com_graph}
\end{figure}
\end{comment}
\begin{figure}[t]
    \centering
    
    \begin{subfigure}{0.20\textwidth}
        \centering
        \includegraphics[width=\linewidth]{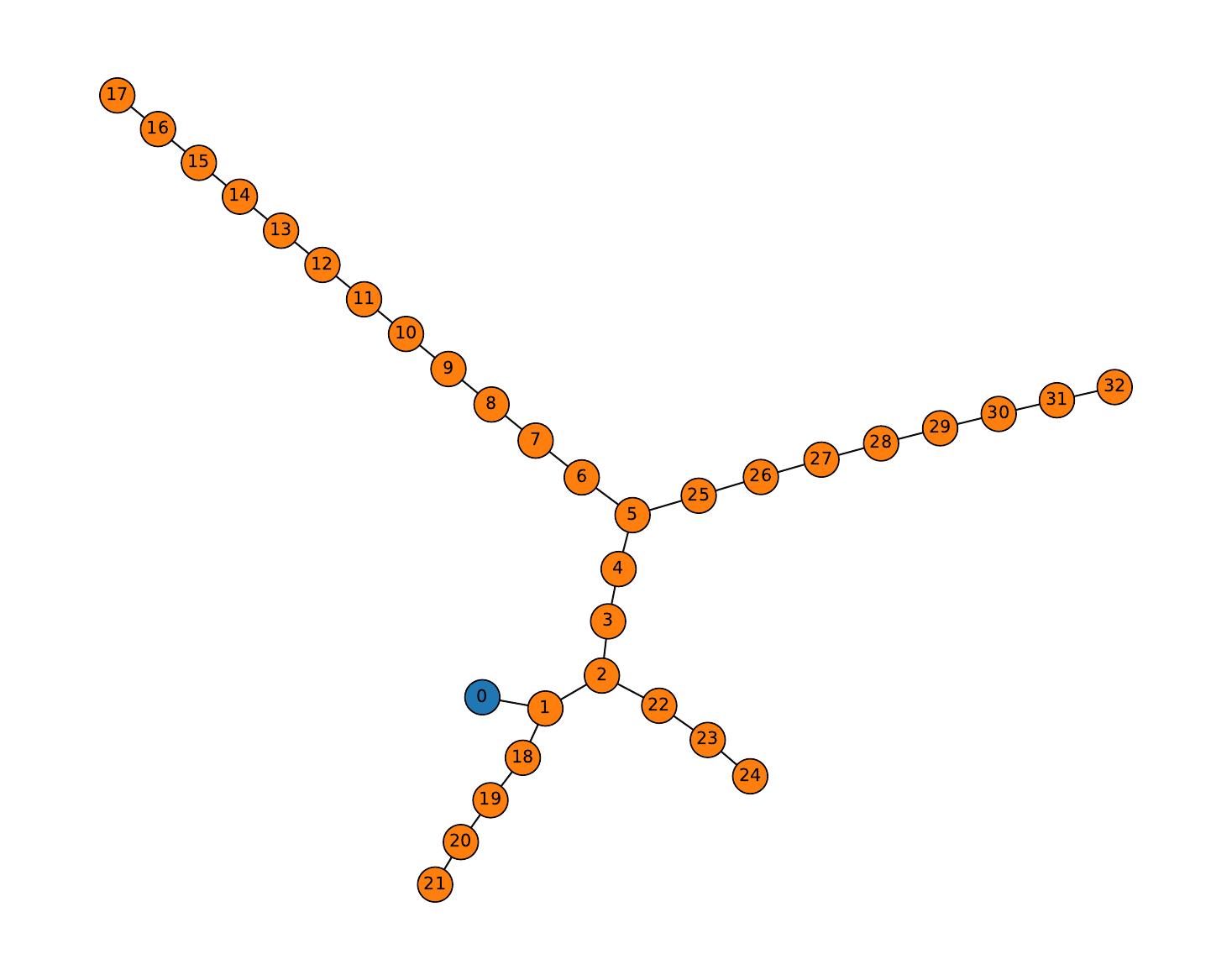}
        \caption{}
    \end{subfigure}
    \hfill
    \begin{subfigure}{0.20\textwidth}
        \centering
        \includegraphics[width=\linewidth]{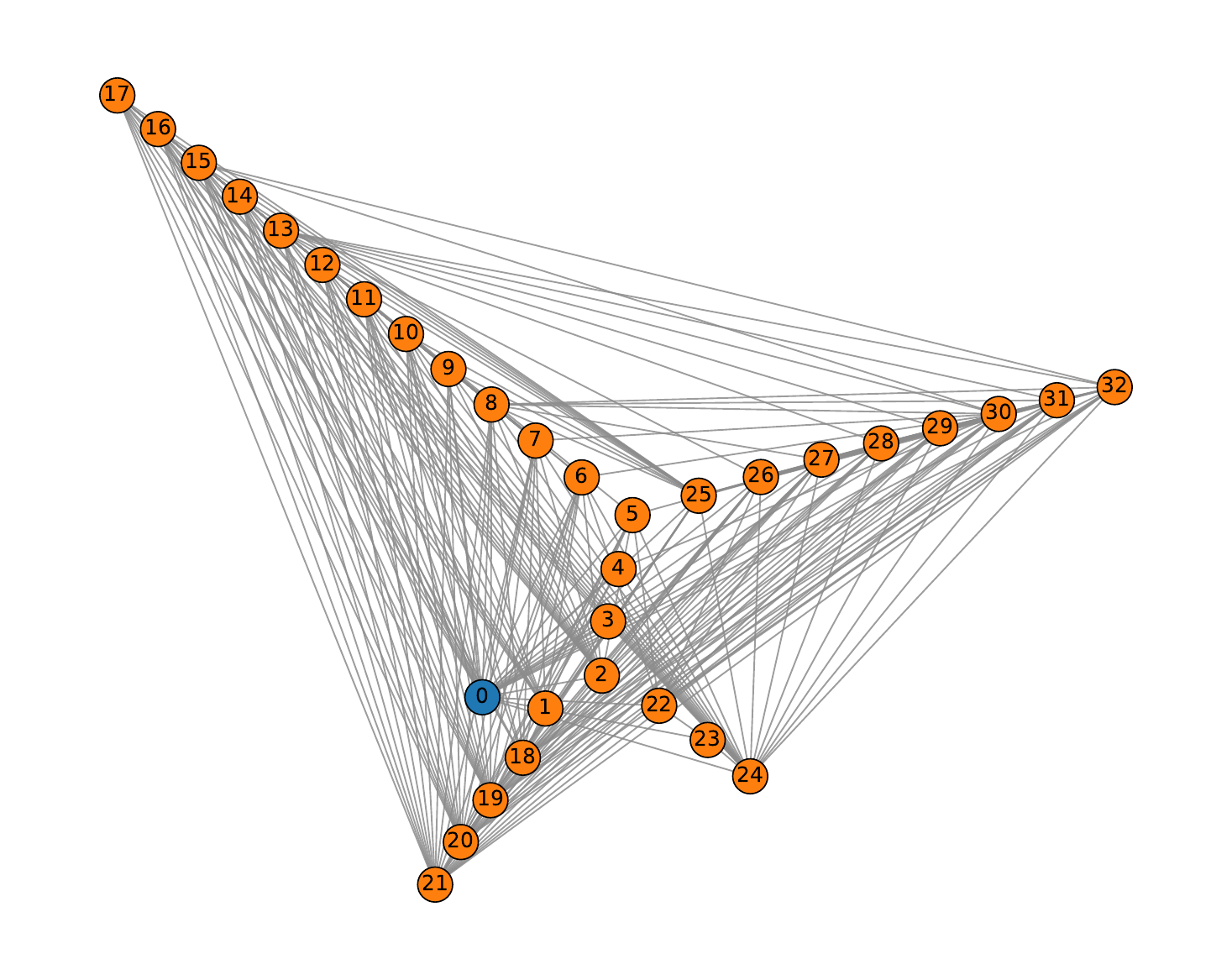}
        \caption{}
    \end{subfigure}

    \vspace{0.1cm}

    \begin{subfigure}{0.20\textwidth}
        \centering
        \includegraphics[width=\linewidth]{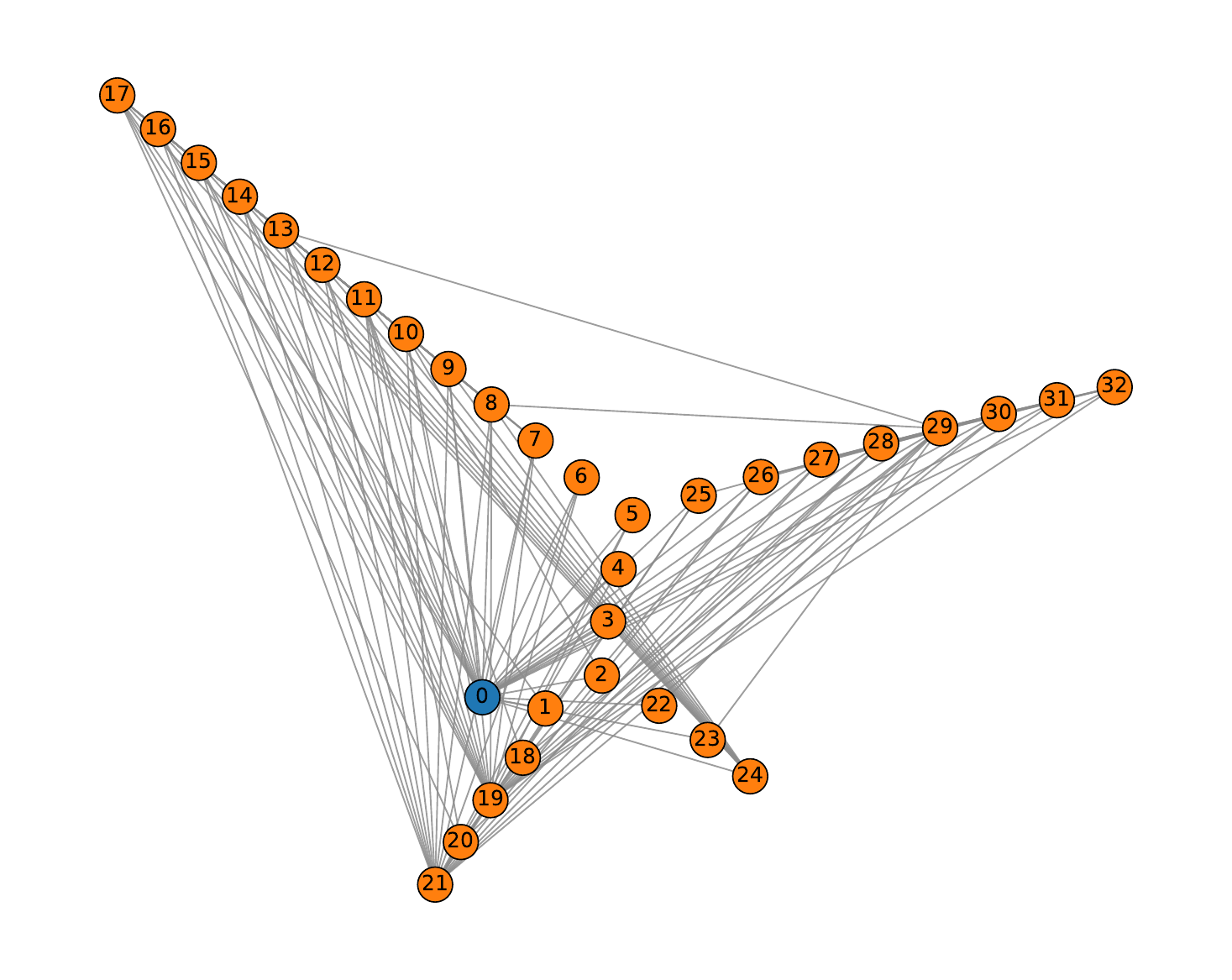}
        \caption{}
    \end{subfigure}
    \hfill
    \begin{subfigure}{0.20\textwidth}
        \centering
        \includegraphics[width=\linewidth]{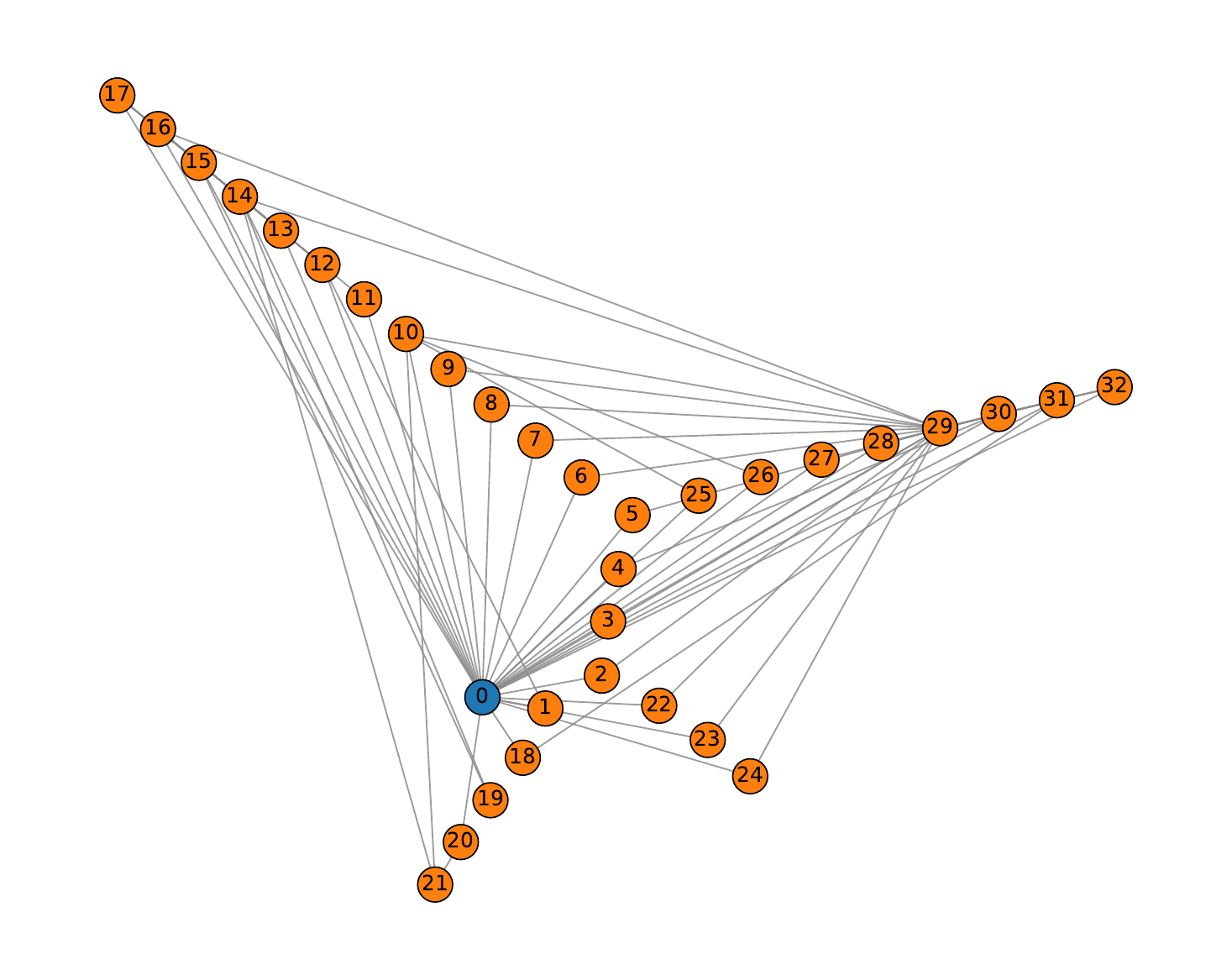}
        \caption{}
    \end{subfigure}

    \caption{(a) Substrate electrical graph, and communication graphs obtained with $\nu_1 = 1.78\cdot10^{-7}$ and (b) $\nu_2 = 1\cdot10^{-6}$, (c) $\nu_2 = 2.51\cdot10^{-6}$, (d) $\nu_2 = 1\cdot10^{-5}$, corresponding to $275$, $148$ and $75$ edges respectively.}
    \label{fig:electr_and_com_graph}
\end{figure}

To illustrate the resulting communication graph for the considered use case, Fig.~\ref{fig:electr_and_com_graph} presents an example obtained using the proposed algorithm, alongside the underlying electrical network. The graphs are obtained by varying $\nu_2$ while holding $\nu_1$ constant, illustrating how the penalization parameter affects the resulting communication topology.
The slack node (blue) connects to the upstream/transmission grid, and consequently has significantly larger capacity than the other resources. This explains why its degree is considerably higher than that of the other nodes. 

\subsection{Empirical Evaluation of the Consensus Algorithm} \label{subsec:Cons-Numerical-Ev}

This subsection evaluates the proposed scheme and compares it to a voltage-agnostic counterpart. The results fall into two categories. The first concerns the \emph{real} voltages, obtained by running a power flow on the dispatch decided at each iteration; the second concerns the algorithm's convergence, reported through the relative duality gap and the cost error induced by neglecting line losses. All runs terminate at the iteration budget $k_{\mathrm{max}} = 2000$ of Table~\ref{tab:params} and the corresponding elapsed time is reported on the x-axis.
The evaluated communication graphs (identified by their number of edges) correspond to the row of Fig.~\ref{fig:Errors} that minimizes the expected RMSE, as described in Subsection~\ref{subsec:UseCase}. 
%For readability, they are identified by their number of edges rather than by the pair $(\nu_1, \nu_2)$.

\begin{figure}[t]
    \centering
    \includegraphics[width=0.5\linewidth]{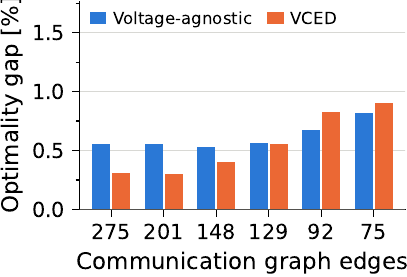}
    \caption{Optimality gap of the voltage-agnostic dispatch and of VCED with respect to the centralised SOCP optimum.}
    %, for communication graphs of decreasing density.}% Gaps are expressed as a fraction of the dispatch-dependent cost.}
    \label{fig:gap_bars}
\end{figure}

\begin{comment}
\begin{figure}[t]
    \centering
    
    \begin{subfigure}{0.24\textwidth}
        \centering
        \includegraphics[width=\linewidth]{images/Voltage_plts/0_Low_DERs/Snapshots_final/vhist_grid_ncols_1_v0.pdf}
    \end{subfigure}
    \hfill
    \begin{subfigure}{0.24\textwidth}
        \centering
        \includegraphics[width=\linewidth]{images/Voltage_plts/0_Low_DERs/Snapshots_final/vhist_grid_ncols_1_v1.pdf}
    \end{subfigure}
    \caption{Distribution of nodal voltages across all buses and time steps at termination (2000 iterations),  in the voltage-agnostic (a) and voltage-aware (b) cases. Dashed lines mark the 0.95–1.05 p.u. range; red bars fall outside it.}
    \label{fig:snapshots1}
\end{figure}    
\end{comment}

\begin{figure}[t]
    \centering
    
    \begin{subfigure}{0.24\textwidth}
        \centering
        \includegraphics[width=\linewidth]{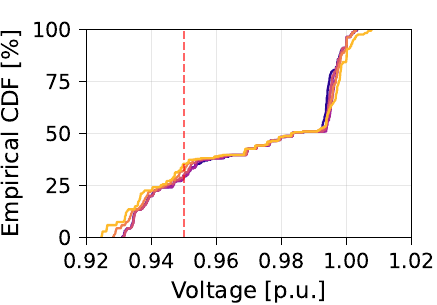}
        \caption{}
    \end{subfigure}
    \hfill
    \begin{subfigure}{0.24\textwidth}
        \centering
        \includegraphics[width=\linewidth]{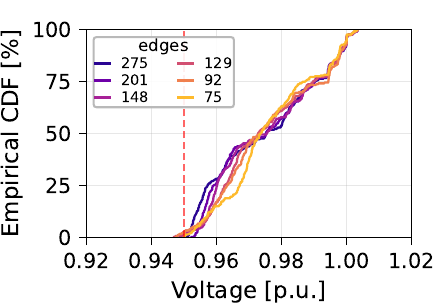}
        \caption{}
    \end{subfigure}
    \caption{Low-DERs scenario Empirical CDF of nodal voltages across all buses and time steps at termination (2000 iterations) for 6 different communication graphs (75--275 edges), in the voltage-agnostic (a) and voltage-aware (b) cases. The dashed line marks the $0.95$~p.u. lower limit, so each curve's height there gives the fraction of undervoltage cases.}
    \label{fig:snapshots2}
\end{figure} 
%\todo[inline]{Pick Fig. \ref{fig:snapshots1} or \ref{fig:snapshots2}}

%Fig.~\ref{fig:snapshots2} shows the corresponding cumulative distribution for the various communication graph configurations. 

We first demonstrate two snapshots of the voltage-agnostic and VCED schemes. Fig.~\ref{fig:gap_bars} reports the optimality gap of both schemes against the centralized OPF-SOCP benchmark, for communication graphs from $275$ down to $75$ edges; %Fig.~\ref{fig:snapshots1} and 
Fig.~\ref{fig:snapshots2} shows the corresponding voltage distributions across all nodes and time steps. While both schemes stay within $1\%$ of the cost of the centralized optimal solution throughout, the voltage-agnostic scheme results in voltage limit violations in more than 25\% of the bus-timesteps.
%, but their ordering depends on connectivity, and therefore accuracy. On the four densest graphs, VCED is the cheaper of the two despite solving a more constrained problem: the voltage term suppresses the high-current conditions that generate losses, and the saving on imported losses outweighs the cost of honoring the band. Below roughly $130$ edges the ordering reverses, as VCED may be too conservative on the voltage estimates in addition to the power balance and is the more sensitive to graph quality; its gap grows from $0.30\%$ to $0.90\%$, against $0.56\%$ to $0.82\%$ for the voltage-agnostic dispatch.
The proposed scheme, in contrast, %is robust. In all communication graphs, our scheme 
produces a clear reduction in the number of bus-timestep pairs lying outside voltage limits, as demonstrated in Fig.~\ref{fig:snapshots2}(b). The violations are not eliminated in every case, but are limited to a small fraction of the total ($<2.65\%$). Across the range of communication graphs considered, from 75 to 275 edges, the resulting voltage distributions are very similar, indicating that the scheme's performance is only weakly affected by the sparsity of the communication graph. This is consistent with Fig.~\ref{fig:Errors}(a), where the expected RMSE of the voltage estimator varies only mildly along this row.

The origin of the persisting voltage violations is worth interpreting. 
%In the classic economic dispatch formulation, decisions are driven purely by economic objectives, without regard for network constraints or quality-of-service requirements such as the admissible voltage range. 
In our approach, nodes operating outside the admissible voltage range are penalized and therefore economically discouraged from ignoring these requirements. However, the penalty acts on the \emph{estimated} voltage rather than the true one, so the residual violations are limited by the accuracy of the surrogate estimator; this is why a small fraction of violations persists even under voltage deviation penalization. We study the contribution of the voltage deviation penalty parameter $\rho^\texttt{v}$ to this effect in Subsection~\ref{subsec:SensitivityAnalysis}.

\begin{figure}[t]
    \centering
    \begin{subfigure}{0.24\textwidth}
        \centering
    \includegraphics[width=\linewidth]{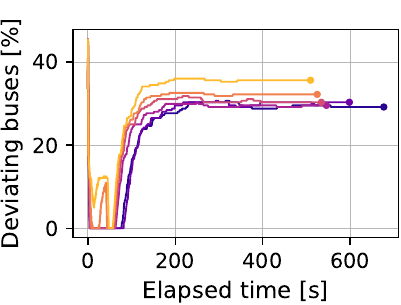}
    \caption{}
    \end{subfigure}
    \hfill
    \begin{subfigure}{0.24\textwidth}
        \centering
    \includegraphics[width=\linewidth]{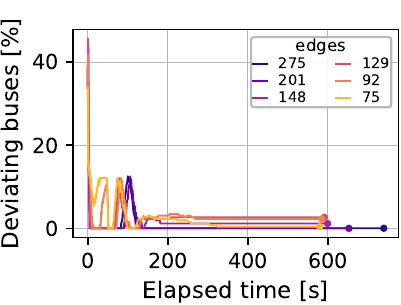}
    \caption{}
    \end{subfigure}    
    \caption{Percentage of bus-timestep voltage violations versus solver time, for the whole community under various communication graphs, in the voltage-agnostic (a) and voltage-aware (b) cases.}
    \label{fig:voltage_dev_perc}
\end{figure}

We now test the online performance of the proposed scheme, i.e. the case where a decision is implemented \textit{before} the algorithm terminates.
Fig.~\ref{fig:voltage_dev_perc} presents the evolution of the real voltage violations across the algortihm's execution time, in both cases: Fig.~\ref{fig:voltage_dev_perc}(a) Voltage-agnostic, and Fig.~\ref{fig:voltage_dev_perc}(b) Voltage-aware. We obtain these results by running the power flow based on the decisions made at each iteration of the algorithm (simulating the case in which the algorithm is interrupted in that iteration). The slack node imports or exports any power deficit or surplus. Voltage improves substantially across all communication graphs/voltage surrogates.
Also, a running time of \SI{200}{\second}--\SI{300}{\second} is sufficient for the voltage to stabilize, after which it does not change appreciably until an acceptable relative duality gap is reached.

To evaluate the convergence of Algorithm~\ref{alg:ConsensusAlgorithm}, we use
the relative duality gap of \ED. Letting
\begin{equation}
  G_i(x_i; \bm{\lambda}) \coloneqq F_i(x_i) + \sum_{t \in \T}
  \big( \lambda^{\texttt{p}}_t\, p_{i,t}(x_i)
  + \lambda^{\texttt{q}}_t\, q_{i,t}(x_i) \big)
  \label{eq:G_i}
\end{equation}
denote agent $i$'s term in the Lagrangian of \ED,
%with respect to the coupling constraints, 
the relative duality gap at iteration $k$ reads
\begin{equation}
  \Delta^{\mathrm{gap}}[k] = \frac{\big| \sum_{i \in \I} F_i(x_i[k])
  - \sum_{i \in \I} G_i\big(x_i[k]; \bm{\lambda}_i[k]\big) \big|}
  {\big| \sum_{i \in \I} F_i(x_i[k]) \big|},
  \label{eq:duality_gap}
\end{equation}
where $\bm{\lambda}_i[k]$ denotes agent $i$’s own multiplier proxies and the slack node’s cost is evaluated at its realized exchange with the upstream grid. The numerator reduces to the multiplier-weighted power-balance residual, which vanishes at a fixed point (refer to Lemma~\ref{lem:lambda} of Appendix A). As reported in Fig.~\ref{fig:rel_dual_gap}, taking $10^{-1}\,\%$ as acceptable, convergence requires around \SI{500}{\second}, and the rate does not change substantially between the voltage-aware and voltage-agnostic cases.

\begin{figure}[t]
    \centering
    \begin{subfigure}{0.24\textwidth}
        \centering
    \includegraphics[width=\linewidth]{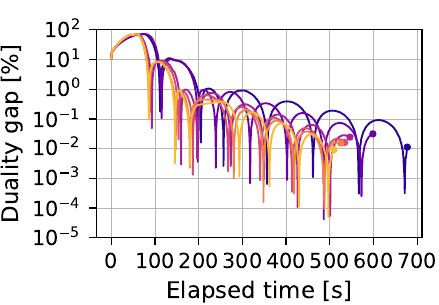}
    \caption{}
    \end{subfigure}
    \hfill
    \begin{subfigure}{0.24\textwidth}
        \centering
    \includegraphics[width=\linewidth]{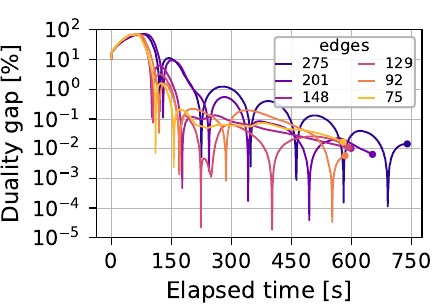}
    \caption{}
    \end{subfigure}    
    \caption{Relative duality gap versus solver time, for the whole community under various communication graphs, in the voltage-agnostic (a) and voltage-aware (b) cases.}
    \label{fig:rel_dual_gap}
\end{figure}

The final test of this subsection relates to the fact that the prosumers do not have access to the network and therefore they cannot account for line losses.
In this test, we evaluate the relative error of the proposed scheme in the community's real cost, that is, the cost that occurs when line losses do apply. 
We compute this by considering the additional energy imported by the slack node at each time step.

\begin{figure}[t]
    \centering
    \begin{subfigure}{0.24\textwidth}
        \centering
    \includegraphics[width=\linewidth]{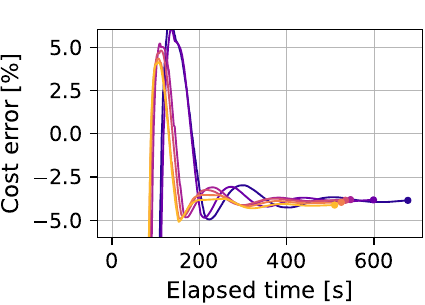}
    \caption{}
    \end{subfigure}
    \hfill
    \begin{subfigure}{0.24\textwidth}
        \centering
    \includegraphics[width=\linewidth]{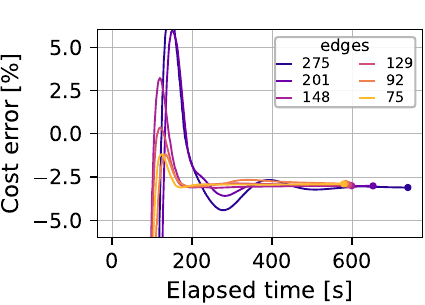}
    \caption{}
    \end{subfigure}    
    \caption{Cost error for the whole community under various communication graphs, in the voltage-agnostic (a) and voltage-aware (b) cases.}
    \label{fig:cost_underest}
\end{figure}

The cost underestimation in Fig.~\ref{fig:cost_underest} settles near $3.5\%$ of the realized cost in the voltage-agnostic case and near $2.5\%$ in the voltage-aware one, in both cases independently of the communication graph, as expected. The reduction is a by-product of voltage regulation: enforcing the band drives the inverters and compensators to higher reactive output, which lowers branch currents and hence losses. The persisting cost error is a consequence of the lossless power balance imposed by the coupling constraint. 
%Both figures are specific to this feeder, and different line impedances or loading levels would change them, so the quantity should be evaluated per system rather than carried over.

\subsection{Sensitivity Analysis}\label{subsec:SensitivityAnalysis}
In this last subsection, we provide a sensitivity analysis with respect to two factors: higher DER at the edges of the network using the voltage surrogate trained under the low-DER scenario, and the value of the penalty weight $\rhv$, which is taken to be common to all agents.

\begin{figure}[t]
    \centering
    
    \begin{subfigure}{0.24\textwidth}
        \centering
        \includegraphics[width=\linewidth]{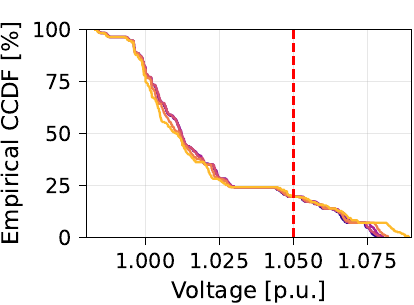}
        \caption{}
    \end{subfigure}
    \hfill
    \begin{subfigure}{0.24\textwidth}
        \centering
        \includegraphics[width=\linewidth]{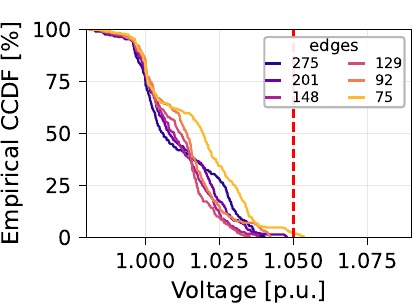}
        \caption{}
    \end{subfigure}
    \caption{High-DERs scenario Empirical CCDF of nodal voltages across all buses and time steps at termination (2000 iterations) for 6 different communication graphs (75--275 edges), in the voltage-agnostic (a) and voltage-aware (b) cases. The dashed line marks the $1.05$~p.u. upper limit, so each curve's height there gives the fraction of overvoltage cases.}
    \label{fig:ecdf_high_der}
\end{figure}

In Fig.~\ref{fig:ecdf_high_der}, all communication graphs eliminate a significant amount of voltage deviations, reported as overvoltage in the voltage-agnostic case, caused by increased power injection at the feeder ends. However, the sparsest communication graph results in a small number of voltage violations, due to poorer voltage estimation performance compared with the rest .

\begin{figure}[t]
    \centering
    
    \begin{subfigure}[t]{0.24\textwidth}
        \centering
        \includegraphics[width=\linewidth]{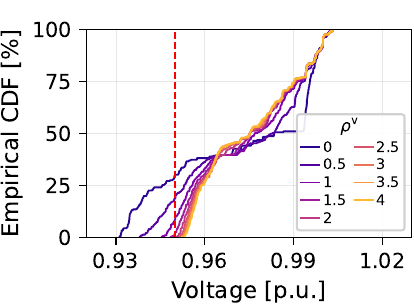}
        \caption{}
    \end{subfigure}
    \hfill
    \begin{subfigure}[t]{0.24\textwidth}
        \centering
        \includegraphics[width=\linewidth]{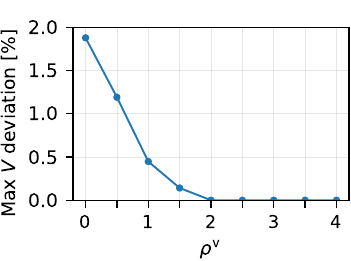}
        \caption{}
    \end{subfigure}
    \caption{Effect of the voltage penalty weight $\rhv$ at termination (2000 iterations), for the 201-edge communication graph: (a) empirical CDF of nodal voltages across all buses and time steps, with the dashed line marking the $0.95$~p.u. lower limit; (b) worst voltage deviation beyond the admissible voltage range.}
    \label{fig:ecdf_rho_v}
\end{figure}

Finally, we test the sensitivity of the algorithm's performance with respect to the voltage penalty weight $\rhv$.
%under a single communication graph. 
The test is run on the communication graph with $\nu_1 = 1.78 \cdot 10^{-7} , \nu_2 = 1.58 \cdot 10^{-6}$ resulting to $201$ communication edges.
In Fig.~\ref{fig:ecdf_rho_v}, we vary 
$\rhv$ to explore its impact on the percentage of deviated nodes and the magnitude of the deviations. In Fig.~\ref{fig:ecdf_rho_v}(a), we observe that for $\rhv \geq 2$, voltage deviations vanish, as expected. In Fig.~\ref{fig:ecdf_rho_v}(b), we focus on the critical region: the range of 
$\rhv$ over which the number of voltage violations transitions to zero. Alongside the reduction in the percentage of nodes with voltage violations, we observe that the maximum deviation also decreases progressively.

\section{Conclusions and Future Work}\label{Sec:Conclusions}

In this paper, we proposed a fully distributed solution to the VCED problem of a community of prosumers, under no knowledge of the physical power grid. To this end, we proposed a linear voltage estimator, based solely on local measurements, and a voltage-aware consensus algorithm.

Our numerical results show that 
%the voltage, projected on a logarithmic scale, at each node can be estimated linearly, with expected RMSE not exceeding 0.6\%. In particular, by varying the LASSO penalty factors, 
voltage estimation accuracy can be traded against the sparsity of the prosumers' communication graph.
%imposed on the electrical grid, without any prior knowledge of its topology or parameters. 
We further show that using this estimated communication graph to solve the VCED in a consensus-based manner substantially reduces voltage deviations,cutting the share of nodes outside the admissible band from about 30\% to below 3\% across every graph tested. The resulting economic suboptimality averaged over the time horizon is bounded analytically: the bound grows with the voltage penalty weight and the total size of the agents' neighborhoods, and quadratically with the error of the voltage surrogate. The trend predicted by this bound is numerically shown as well.
%Overall, this work contributes to the field of distributed, consensus-based optimal power flow and economic dispatch by showing that voltage-aware coordination can be achieved without prior knowledge of the grid's topology or electrical parameters, using a learned communication structure, across a wide range of sparsity levels, in place of the physical network model. This offers a practical path toward voltage-aware distributed dispatch in settings where full grid model information is unavailable, proprietary, or costly to obtain, while still achieving near-optimal cost performance. 
%It is worth noting, however, that this work assumes the availability of a sufficient set of historical or simulated voltage and power measurements to train the estimator, as well as lossless lines in the power balance constraints, which may introduce some inconsistency between the reported dispatch cost and the true cost incurred in real-world, lossy grids. Furthermore, improving the accuracy of the voltage estimator—and consequently reducing the displacement of the optimal dispatch point from the true cost optimum—could enable even sparser communication graphs.

Building on these results, several directions remain open for future research. First, relaxing the lossless-line assumption by incorporating power losses into the VCED formulation would improve the consistency between the reported and true dispatch costs in real-world grids. Second, evaluating the scalability of the proposed estimator and consensus-based scheme on larger, real-world distribution networks would help assess its practical applicability beyond the test cases considered in this work. Also, designing an adaptive or online version of the voltage estimator that can update the communication graph in response to time-varying grid conditions or topology changes, without requiring full retraining, would be valuable. Additionally, it would be worthwhile to study the estimator's robustness to noisy, missing, or adversarially corrupted measurements, which are common in practical deployments. Finally, a natural extension would be to further improve estimation accuracy to reduce the displacement of the consensus-based optimal dispatch point from the true cost optimum, potentially enabling even sparser communication graphs without sacrificing performance.

\section*{Acknowledgment}
The authors would like to thank J.~F.~Heredia of UPC and E.~J.~Salazar of IREC for helpful discussions.

\appendices
\section{Proof of Theorem~\ref{theor:optimality}}
\label{Sec:Apndx_A}
 
In this appendix we define the fixed points of Algorithm~\ref{alg:ConsensusAlgorithm} and then prove, in order, Lemma~\ref{lem:conserv}, Lemma~\ref{lem:lambda} and Theorem~\ref{theor:optimality}.

Throughout, fix $t \in \T$ and collect the agents' quantities in the vectors
$\bm{\lambda}^{\texttt{p}}_t[k], \bm{\delta}^{\texttt{p}}_t[k],
\mathbf{p}_t[k] \in \mathbb{R}^{|\I|}$, with $\mathbf{W} = [W_{i,j}]$ and
$W_{i,j} = 0$ for $(i,j) \notin \E$. Updates \eqref{eq:Plambda_update} and
\eqref{eq:ResidualP} then read
\begin{align}
  \bm{\lambda}^{\texttt{p}}_t[k] &= \mathbf{W}\bm{\lambda}^{\texttt{p}}_t[k-1]
  + \epsp\, \bm{\delta}^{\texttt{p}}_t[k-1], \label{eq:lambda_vec}\\
  \bm{\delta}^{\texttt{p}}_t[k] &= \mathbf{W}\bm{\delta}^{\texttt{p}}_t[k-1]
  + \mathbf{p}_t[k] - \mathbf{p}_t[k-1]. \label{eq:delta_vec}
\end{align}
The reactive quantities evolve identically with $\texttt{p}$ replaced by
$\texttt{q}$; all results hold for both and are stated for active power only.
Assumption~\ref{ass:convex} holds for the models of Section~\ref{Sec:Problem_Form}: each $f_i$ is convex and the local constraints are affine or second-order cone constraints.

Seen as a function of agent $i$'s decision, \eqref{eq:ResidualP} reads $\ep[k] = \ep(x_i[k])$, and line~4 of Algorithm~\ref{alg:ConsensusAlgorithm} reads
$x_i[k] \in \operatorname*{arg\,min}_{x_i \in \Omega_i} \mathcal{L}_i(x_i)$,
with $\mathcal{L}_i$ the local augmented Lagrangian
\eqref{eq:aug_lagrang_indiv} written through Eq.~\eqref{eq:ResidualP}. Note
that $p_{i,t}$ is itself a component of $x_i$, by
\eqref{eq:agent_active_power_balance}.

We call $(x^\star, \bm{\lambda}^\star, \bm{\delta}^\star)$ a \emph{fixed
point} of Algorithm~\ref{alg:ConsensusAlgorithm} if one further iteration
reproduces it. Since then
$\mathbf{p}_t[k] = \mathbf{p}_t[k-1] = \mathbf{p}^\star_t$, this is equivalent
to
\begin{align}
  \bm{\lambda}^{\texttt{p}\star}_t &= \mathbf{W}\bm{\lambda}^{\texttt{p}\star}_t
  + \epsp\, \bm{\delta}^{\texttt{p}\star}_t, \label{eq:fp_lambda}\\
  \bm{\delta}^{\texttt{p}\star}_t &= \mathbf{W}\bm{\delta}^{\texttt{p}\star}_t,
  \label{eq:fp_delta}
\end{align}
together with each $x_i^\star$ minimizing $\mathcal{L}_i$ over $\Omega_i$ at $\bm{\lambda}^\star, \bm{\delta}^\star, \mathbf{p}^\star$. 
%The results below apply verbatim to the limit of any convergent sequence of iterates, since the updates are linear and the first-order conditions of the local problems are preserved in the limit.
The limit of any convergent sequence of iterates is itself a fixed point, by continuity of the updates and Assumption~\ref{ass:convex}, so the results below apply to it unchanged.

\begin{lemma}[Conservation of the aggregate residual]\label{lem:conserv}    
Let Assumptions~\ref{ass:graph} and~\ref{ass:init} hold. Then, at every
iteration $k \geq 0$ and for every $t \in \T$,
$\sum_{i \in \I} \ep[k] = \sum_{i \in \I} \p[i][k]$ and $  \sum_{i \in \I} \eq[k] = \sum_{i \in \I} \q[i][k]$. That is, the local residual estimates are collectively truthful: their sum
reproduces the true community imbalance, although no single agent observes it.
\end{lemma}
\begin{IEEEproof}[Proof of Lemma~\ref{lem:conserv}]
By induction on $k$. For $k = 0$ the claim is
Assumption~\ref{ass:init}. Assume it holds at $k-1$. Premultiplying
\eqref{eq:delta_vec} by $\mathbf{1}^\top$ and using
$\mathbf{1}^\top\mathbf{W} = \mathbf{1}^\top$ gives
\[
  \mathbf{1}^\top \bm{\delta}^{\texttt{p}}_t[k]
  = \mathbf{1}^\top \bm{\delta}^{\texttt{p}}_t[k-1]
  + \mathbf{1}^\top \mathbf{p}_t[k] - \mathbf{1}^\top \mathbf{p}_t[k-1]
  = \mathbf{1}^\top \mathbf{p}_t[k].
\]
Only column stochasticity is necessary here.
\end{IEEEproof}

\vspace{4pt}

\begin{lemma}[Characterization of fixed points]\label{lem:lambda}
Let Assumptions~\ref{ass:graph} and~\ref{ass:init} hold and let
$(x^\star, \bm{\lambda}^\star, \bm{\delta}^\star)$ be a fixed point of
Algorithm~\ref{alg:ConsensusAlgorithm}. Then, for every $t \in \T$, the residual
estimates vanish, and the multipliers reach consensus,
\[
  \bm{\delta}^{\texttt{p}\star}_t = \bm{\delta}^{\texttt{q}\star}_t
  = \mathbf{0}, \qquad
  \bm{\lambda}^{\texttt{p}\star}_t = \bar{\lambda}^{\texttt{p}}_t \mathbf{1},
  \qquad
  \bm{\lambda}^{\texttt{q}\star}_t = \bar{\lambda}^{\texttt{q}}_t \mathbf{1},
\]
for some $\bar{\lambda}^{\texttt{p}}_t, \bar{\lambda}^{\texttt{q}}_t \in
\mathbb{R}$, and the community power balance
\eqref{eq:active_power_balance}--\eqref{eq:reactive_power_balance} holds at
$x^\star$.
\end{lemma}
\begin{IEEEproof}[Proof of Lemma~\ref{lem:lambda}]

Since $\graph$ is connected, $\mathbf{W}$ is irreducible. By the Perron--Frobenius theorem~\cite{frobenius1912}, \cite[Ch.~8]{horn2012matrix}, the eigenvalue $1$ of the irreducible stochastic matrix $\mathbf{W}$ is simple with eigenvector $\mathbf{1}$, so $\ker(\mathbf{I}-\mathbf{W}) = \operatorname{span}\{\mathbf{1}\}$. Hence \eqref{eq:fp_delta} gives $\bm{\delta}^{\texttt{p}\star}_t = c\,\mathbf{1}$ for some $c \in \mathbb{R}$. Premultiplying \eqref{eq:fp_lambda} by $\mathbf{1}^\top$ and using $\mathbf{1}^\top\mathbf{W} = \mathbf{1}^\top$ yields $0 = \epsp\, c\, |\I|$, so $c = 0$ because $\epsp > 0$ and thus $\bm{\delta}^{\texttt{p}\star}_t = \mathbf{0}$. Equation \eqref{eq:fp_lambda} then reduces to $\bm{\lambda}^{\texttt{p}\star}_t = \mathbf{W}\bm{\lambda}^{\texttt{p}\star}_t$, and the same eigenvalue argument gives $\bm{\lambda}^{\texttt{p}\star}_t = \bar{\lambda}^{\texttt{p}}_t\mathbf{1}$. Finally, $\bm{\delta}^{\texttt{p}\star}_t = \mathbf{0}$ and Lemma~\ref{lem:conserv} give $\sum_{i \in \I} p^\star_{i,t} = 0$, i.e.\ \eqref{eq:active_power_balance}.
\end{IEEEproof}

\vspace{4pt}

\begin{IEEEproof}[Proof of Theorem~\ref{theor:optimality}]
Throughout, $F_i$ is as in Assumption~\ref{ass:convex} and $G_i$ as in \eqref{eq:G_i}, evaluated at the fixed-point multipliers $\bar{\bm{\lambda}}$.

\emph{Step 1 (local optimality).} By \eqref{eq:ResidualP}, $\nabla \ep(x_i) = \nabla p_{i,t}$, a constant unit vector since $p_{i,t}$ is a component of $x_i$. By \eqref{eq:ResidualP} and Lemma~\ref{lem:lambda}, $\ep(x_i^\star) = (\mathbf{W}\bm{\delta}^{\texttt{p}\star}_t)_i = 0$ and $\lambda^{\texttt{p}\star}_{i,t} = \bar{\lambda}^{\texttt{p}}_t$, and likewise for the reactive quantities. Differentiating
\eqref{eq:aug_lagrang_indiv} therefore gives
\begin{multline*}
  \nabla \mathcal{L}_i(x_i^\star) = \nabla F_i(x_i^\star)
  + \sum_{t \in \T} \Big[ \big(\bar{\lambda}^{\texttt{p}}_t
  + \rhop\, \ep(x_i^\star)\big) \nabla p_{i,t} \\
  + \big(\bar{\lambda}^{\texttt{q}}_t
  + \rhoq\, \eq(x_i^\star)\big) \nabla q_{i,t} \Big]
  = \nabla G_i(x_i^\star).
\end{multline*}
Since $x_i^\star$ minimizes the convex function $\mathcal{L}_i$ over $\Omega_i$, the first-order condition $-\nabla \mathcal{L}_i(x_i^\star) \in \mathcal{N}_{\Omega_i}(x_i^\star)$ holds, with $\mathcal{N}_{\Omega_i}$ the normal cone of $\Omega_i$. Hence $-\nabla G_i(x_i^\star) \in \mathcal{N}_{\Omega_i}(x_i^\star)$ and, $G_i$ being convex,
\begin{equation}
  x_i^\star \in \operatorname*{arg\,min}_{x_i \in \Omega_i} G_i(x_i),
  \quad \forall i \in \I.
  \label{eq:lagrangian_min}
\end{equation}

\emph{Step 2 (global optimality).} By Lemma~\ref{lem:lambda}, $x^\star$ is feasible for \ED. Let $x$ be any other feasible point. Since the coupling constraints hold at both $x$ and $x^\star$, the linear terms of \eqref{eq:G_i} vanish in aggregate, and \eqref{eq:lagrangian_min} gives
\[
  \sum_{i \in \I} F_i(x_i) = \sum_{i \in \I} G_i(x_i)
  \geq \sum_{i \in \I} G_i(x_i^\star) = \sum_{i \in \I} F_i(x_i^\star),
\]
so $x^\star$ is optimal for \ED. Moreover $x^\star$ minimizes the Lagrangian $\sum_{i} G_i$ over $\prod_{i} \Omega_i$ and satisfies the coupling constraints, so $(x^\star, \bar{\bm{\lambda}}^{\texttt{p}}, \bar{\bm{\lambda}}^{\texttt{q}})$ is a saddle point of the Lagrangian of \ED. The multipliers are therefore optimal, with zero duality gap.
\end{IEEEproof}

\vspace{4pt}

\begin{remark}[Role of the augmented terms]\label{rem:penalty}
The quadratic terms of \eqref{eq:aug_lagrang_indiv} act on the residual estimates, which vanish at a fixed point by Lemma~\ref{lem:lambda}, so their contribution to the optimality condition vanishes: they shape the algorithm's trajectory but not its fixed points. This is because they add curvature to $F_i$, which increases solution stability, especially for agents whose costs are zero or linear.
\end{remark}

\begin{remark}[Uniqueness]\label{rem:uniqueness}
The curvature added by the quadratic terms renders each local problem strictly convex in the directions affecting the agent's net injections, so the local minimizers are unique. Problem \ED may nevertheless admit several optima, in which case different fixed points correspond to different ones, each of them optimal.
\end{remark}

\begin{remark}[Convergence]\label{rem:convergence}
The results above characterize the fixed points of Algorithm~\ref{alg:ConsensusAlgorithm}; neither their existence nor the convergence of the iterates to one of them is shown here. Step-size conditions guaranteeing convergence are available for the broader family of augmented Lagrangian and dual-ascent methods~\cite{Betsekas_Constrained_Opt_1982} and, for voltage-agnostic economic dispatch, in~\cite{Wu_2026}. These do not transfer directly to the present scheme, which dualizes two coupling constraints and includes the voltage penalty \eqref{eq:F_indiv_voltage} in each local objective, but they indicate the order of magnitude of admissible dual step sizes; we tuned the values used in Section~\ref{Sec:Results} accordingly. 
%Note that these conditions concern convergence to a fixed point, not its relation to \ED, which is the object of Theorem~\ref{theor:optimality}.
\end{remark}

\section{Proof of Theorem~\ref{theor:error_quant}}
\label{Sec:Apndx_B}

\begin{IEEEproof}
Let $h(v) := \sum_{i\in \I, t \in \T} \Ni h_i(v_{i,t})$, where $ h_i(v_{i,t})= [(v_{i,t} - \overline{V})_+^2 + (\underline{V} - v_{i,t})_+^2]$ and  $F(x) \coloneqq \sum_{i \in\I} F_i(x_{i})$. Let $\xa$ and $\xb$ denote the global minimizers of the problem solved by Algorithm~\ref{alg:ConsensusAlgorithm} and of (VCED), respectively. Equivalently, $\xb$ is the minimizer obtained by Algorithm~\ref{alg:ConsensusAlgorithm} when the exact voltage mapping $\Phi(\cdot)$ is used and the voltage penalties $\rhv$ and $\rho^{\delta\texttt{v}}$ are sufficiently large.
Considering Assumption~\ref{ass:volt_cons}, we have
\begin{align}
    F(\xa) + f(\Phi(\xa)) \geq F(\xb) + f(\Phi(\xb)) \label{eq:relax_optimal}\\
    F(\xa) + f(\phi(\xa)) \leq F(\xb) + f(\phi(\xb)) \label{eq:strict_optimal} 
\end{align}
As $h(\Phi(\xb)) = 0$, by definition, and considering Eq.~\eqref{eq:relax_optimal}, \eqref{eq:strict_optimal} we get:

\begin{equation}
    F(\xa) - F(\xb) \leq h(\phi(\xb))-h(\phi(\xa)) \label{ineq:costs}
\end{equation}

Without the loss of generality, we can evaluate the upper bound of Eq.~\eqref{ineq:costs} for the case where mappings $\Phi(x), \phi(x)$ lie close to the upper voltage bound and consider that:

\begin{equation}
    H_{min}(\phi(x)) \leq h(\phi(x))\leq H_{max}(\phi(x))
\end{equation}
we bound from above the right-hand side of Eq.~\eqref{ineq:costs} and consequently the cost difference as below:

\begin{equation}
  - H_{min}(\Phi(\xa)) \leq F(\xa) - F(\xb) \leq H_{max}(\phi(\xb)) - H_{min}(\phi(\xa)) \label{ineq:cost_bound}
\end{equation}

We set $\distPhi(x)= \Phi_i(x_t) - \overline{V}$, $y_+ = \max(y,0)$ and $y_- = \min(y,0)$. Considering Assumption~\ref{ass:sur_acc} we have that:

\begin{align}
    H_{min}(\phi(x)) = \rhv\sum_{i\in \I, t \in \T}&[\Ni (\distPhi(x_t) - \epsv )_+^2] \notag \\
     = \rhv\sum_{i\in \I, t \in \T}\Ni& [\distPhi(x_t)_+^2 -2 \epsv \distPhi(x_t)_+ \notag\\ &+ \epsv^2 -(\epsv - \distPhi(x_t)_+)_+^2] \notag\\
     \geq \rhv\sum_{i\in \I, t \in \T}\Ni& [\distPhi(x_t)_+^2 -2 \epsv \distPhi(x_t)_+]
 \end{align}

\begin{align}
    H_{max}(\phi(x)) = \rhv\sum_{i\in \I, t \in \T}&\Ni[ \distPhi(x_t) + \epsv )_+^2] \notag\\
     = \rhv\sum_{i\in \I, t \in \T}\Ni& [\distPhi(x_t)_+^2 + 2 \epsv \distPhi(x_t)_+ \notag\\ & +(\epsv + \distPhi(x_t)_-)_+^2] \notag\\
     \leq \rhv\sum_{i\in \I, t \in \T}\Ni& [\distPhi(x_t)_+^2 +2 \epsv \distPhi(x_t)_+ + \epsv^2]
 \end{align}

Considering Eq.~\eqref{ineq:cost_bound} and $H_{min}, H_{max}$ bounds we get:
\begin{align}
 H_{max}(&\phi(\xb)) - H_{min}(\phi(\xa))
  \le \rhv \sum_{i\in\I,\, t\in\T} \Ni \Bigl[ \distPhi(\xb_t)_+^2   \notag\\
  &\phantom{{}wrt}
     {}+ 2\epsv\, \distPhi(\xb_t)_+ + \epsv^2 - \distPhi(\xa_t)_+^2 + 2\epsv\, \distPhi(\xa_t)_+ \Bigr] \notag\\
  &= \rhv \sum_{i\in\I,\, t\in\T} \Ni \Bigl[ \epsv^2 + 2\epsv\, \distPhi(\xa_t)_+ - \distPhi(\xa_t)_+^2  \Bigr]\notag \\
  & \leq 2\rhv |\T| \Gamma \epsv^2 \label{ineq:cost_bound_2}
\end{align}

%Where parameters are $A =\rhv |\T|\sum_{i\in\I} \Ni$, $S_1 = 2 \rhv \sum_{i\in\I,\, t\in\T}  \distPhi(\xa_t)_+$ and $S_2 =  \rhv \sum_{i\in\I,\, t\in\T} \distPhi(\xa_t)_+^2$ 

\begin{comment}

Each slack enters \eqref{LocalProblem} only through its own bound \eqref{eq:UpSlack}--\eqref{eq:DownSlack} and its square in \eqref{eq:F_indiv_voltage}, so at any minimizer $\UpSlackVoltage = (\Vijt - \overline{V})_+$ and $\DownSlackVoltage = (\underline{V} - \Vijt)_+$, positive only where the estimate leaves the band. With $\Vijt = \phi_j$ and the disagreement term of \eqref{eq:F_indiv_voltage} zero by Assumption~\ref{ass:volt_cons}, $\mathcal{V}_i = \rhv \sum_{j \in \Ni} h(\phi_j)$; according to Theorem~\ref{theor:optimality}, $\xa$ is a minimizer of $\sum_{i \in \I} F_i(x_i) + \rhv \sum_{t \in \T} \sum_{i \in \I} \sum_{j \in \Ni} h(\phi_j)$ over the feasible set of \ED. Since (ED) is \eqref{centralized} without \eqref{eq:voltage_limits}, any minimizer $x^{\texttt{c}}$ of \eqref{centralized} is feasible for \ED, and there $\underline{V} \le \Phi_j \le \overline{V}$. Assumption~\ref{ass:sur_acc} bounds each of $(\phi_j - \overline{V})_+$, $(\underline{V} - \phi_j)_+$ by $\epsv$ with at most one of both being positive, giving $h(\phi_j) \le \epsv^2$ over the $|\T|\Gamma$ terms of the penalty.
\end{comment}
Hence we get that
\begin{equation*}
  \sum_{i \in \I} F_i(\xa_i) \;\le\; \sum_{i \in \I} F_i(\xb_i) + 2\rhv |\T| \Gamma \epsv^2 \;=\; H^\star + 2\rhv |\T| \Gamma \epsv^2 ,
\end{equation*}
by minimality of $\xb$ and non-negativity of $h$ at $\xa$.
\end{IEEEproof}
\section{Communication Matrix Construction} \label{Sec:Apndx_C}
Let $\graph = (\I, \E)$ be the communication graph returned by Algorithm~\ref{alg:graph_construction}. We define the adjacency matrix
$\mathbf{A} \in \{0,1\}^{|\I| \times |\I|}$, the graph Laplacian $\mathbf{L} \in \mathbb{R}^{|\I| \times |\I|}$, and the row-stochastic matrix $\tilde{\mathbf{W}} \in \mathbb{R}^{|\I| \times |\I|}$ as
\begin{align}
  A_{i,j} &= \begin{cases}
    1, & \text{if } j \in \Ni \setminus \{i\},\\
    0, & \text{otherwise,}
  \end{cases} \label{eq:adj_mat}\\
  L_{i,j} &= \begin{cases}
    \sum_{l \in \I} A_{i,l}, & \text{if } j = i,\\
    -A_{i,j}, & \text{otherwise,}
  \end{cases} \label{eq:laplacian_mat}\\
  \tilde{W}_{i,j} &= \frac{|L_{i,j}|}{\sum_{l \in \I} |L_{i,l}|}.
  \label{eq:com_mat}
\end{align}

By construction, $\tilde{\mathbf{W}}$ in \eqref{eq:com_mat} is nonnegative and
row stochastic, i.e., $\tilde{\mathbf{W}}\mathbf{1} = \mathbf{1}$, with
$\tilde{W}_{i,j} > 0$ if and only if $(i,j) \in \E$. However, it is column
stochastic only if all nodes have the same degree, i.e., if $\graph$ is
$k$-regular, which is generally not the case for the graphs produced by
Algorithm~\ref{alg:graph_construction}.

To remove the need for a $k$-regular graph, we apply the Sinkhorn--Knopp
algorithm~\cite{sinkhorn1967concerning}, which alternately normalizes the rows
and columns of $\tilde{\mathbf{W}}$ and converges to a matrix
$\mathbf{W} = \mathbf{D}_1 \tilde{\mathbf{W}} \mathbf{D}_2$, with
$\mathbf{D}_1, \mathbf{D}_2$ positive diagonal, that satisfies
\begin{equation}
  \mathbf{W}\mathbf{1} = \mathbf{1}, \qquad
  \mathbf{1}^\top \mathbf{W} = \mathbf{1}^\top,
\end{equation}
where $\mathbf{1}$ is the all-ones vector. Convergence is guaranteed because $\tilde{\mathbf{W}}$ has total support: its zero pattern is symmetric and its diagonal is positive~\cite{sinkhorn1967concerning}. Since the scaling is diagonal, $\mathbf{W}$ has the same zero pattern as $\tilde{\mathbf{W}}$, so the communication graph is preserved. Moreover, the doubly stochastic diagonal scaling of a matrix is unique~\cite{sinkhorn1967concerning}. Since $\tilde{\mathbf{W}}$ is itself a diagonal scaling of the symmetric matrix $|\mathbf{L}|$, the resulting $\mathbf{W}$ is also symmetric. The iteration terminates when the row and column sums deviate from unity by at most $\varepsilon_{\mathrm{SK}} = 10^{-12}$, which is negligible relative to the convergence tolerances of Algorithm~\ref{alg:ConsensusAlgorithm}.

\bibliographystyle{IEEEtran}
\bibliography{references}

\end{document}